\documentclass[preprint,nofootinbib,a4paper,11pt]{revtex4}

\usepackage[centertags]{amsmath}
\usepackage{amsfonts}
\usepackage{amssymb}
\usepackage[sort&compress]{natbib}
\usepackage[hypertex]{hyperref}

\DeclareMathOperator{\Sp}{Sp}

\DeclareMathOperator{\gh}{gh}

\DeclareMathOperator{\sign}{sign}

\newcommand{\spp}{\mathbf{p}}

\newcommand{\spk}{\mathbf{k}}

\newcommand{\spx}{\mathbf{x}}

\newcommand{\spy}{\mathbf{y}}

\newcommand{\lan}{\langle}

\newcommand{\ran}{\rangle}

\newcommand{\e}{\varepsilon}

\newcommand{\vf}{\varphi}

\newcommand{\s}{\sigma}

\newcommand{\al}{\alpha}

\newcommand{\ga}{\gamma}

\newcommand{\de}{\delta}

\newcommand{\ka}{\varkappa}

\newcommand{\la}{\lambda}

\newcommand{\ups}{\upsilon}

\begin{document}

\title{Fluctuations as stochastic deformation}

\date{\today}

\author{P.O. Kazinski}

\email{kpo@phys.tsu.ru}

\affiliation{Physics Faculty, Tomsk State University, Tomsk, 634050 Russia}

\begin{abstract}

A notion of stochastic deformation is introduced and the corresponding algebraic
deformation procedure is developed. This procedure is analogous to the deformation of an
algebra of observables like deformation quantization, but for an imaginary
deformation parameter (the Planck constant). This method is demonstrated on diverse
relativistic and nonrelativistic models with finite and infinite degrees of freedom. It
is shown that under stochastic deformation the model of a nonrelativistic particle
interacting with the electromagnetic field on a curved background passes into the
stochastic model described by the Fokker-Planck equation with the diffusion tensor being
the inverse metric tensor. The first stochastic correction to the Newton equations for
this system is found. The Klein-Kramers equation is also derived as the stochastic
deformation of a certain classical model. Relativistic generalizations of the
Fokker-Planck and Klein-Kramers equations are obtained by applying the procedure of
stochastic deformation to appropriate relativistic classical models. The analog of the
Fokker-Planck equation associated with the stochastic Lorentz-Dirac equation is derived
too. The stochastic deformation of the models of a free scalar field and an
electromagnetic field is investigated. It turns out that in the latter case the obtained
stochastic model describes a fluctuating electromagnetic field in a transparent medium.

\end{abstract}

\pacs{05.40.-a}


\maketitle

\section{Introduction}

Stochastic equations are usually considered as an effective tool that is able to take into account infinitely many factors acting on an open system. Generally, they are obtained by adding to the classical equations of motion of the system a random force (noise) with some probability distribution law. In spite of this common view we shall develop here a slightly different technique inspired by an algebraic approach to quantum mechanics. As is known quantum fluctuations or quantum mechanics itself can be regarded as resulting from a deformation of underlying classical mechanics or, more precisely, from an algebraic deformation of the Poisson structure on the phase space of the classical system. This approach, known as deformation quantization, was initiated in the
seminal papers \cite{BFFLS}. Since the quantum fluctuations arise from the deformation of the Poisson structure, it is reasonable to pose the question whether the stochastic
fluctuations could be reproduced in such a way. In this paper we give an affirmative
answer on this question and sustain it by diverse examples.

A similarity between equations of quantum and stochastic mechanics was perceived by many
authors starting from Schr\"{o}dinger himself \cite{Shro}. Much afterwards this
similarity was brought in an almost perfect accordance by Zambrini in his works
\cite{Zambr} on Euclidean quantum mechanics\footnote{It is noteworthy to say
that unlike Zambrini's works we do not construct the resulting stochastic mechanics as a
solution of the so-called Schr\"{o}dinger gedanken experiment but expound it in a rather
traditional way.}. However the key point of these works as well as the works of other authors \cite{FeynStat,Nam,Par,Ris,ZJ} was an
analytic continuation to imaginary time and not the deformation of the Poisson structure.
Further we shall see that in the relativistic case this slight disagreement is not a
matter of interpretation but gives inequivalent results.

The principal point of this paper is that we can obtain stochastic mechanics deforming
the associative and commutative product of the algebra of classical observables (the
smooth functions over the phase space) along the same lines as in the deformation
quantization procedure. However contrary to an
ordinary quantum mechanics with the real deformation parameter (the Planck constant) we deform the product by an imaginary deformation parameter.
By analogy we shall refer to this procedure as stochastic deformation\footnote{We do not
use the more familiar term ``stochastic quantization'' by two reasons. Firstly, this
term refers to the fixed notion of theoretical physics which is not concerned with what we are considering in this paper. Secondly, there is nothing quantal in the theory developed
here. That is to say such a deformation does not result in a quantization of observables.}. The deformation parameter, which we denote by $\hbar$, characterizes the ``openness''
of the system and a variance of random forces is proportional to it. The mention should
be made that the Hamiltonian generating the dynamics of stochastic system may take
forms unexpected at first glance, because the classical limit $\hbar\rightarrow0$ of stochastic mechanics is not an ordinary classical mechanics. In this limit we obtain some classical mechanics where the momenta should be
regarded as systematic forces (or proportional to them) acting on the system, for this
limit is equivalent to a treatment of the system in the equilibrium state when stochastic fluctuations vanish and the effect of systematic forces is balanced by a dissipation. The expansion of the $pq$-symbol of the Hamiltonian in momenta is closely related to the
Kramers-Moyal expansion and the coefficients of this expansion are connected with the
cumulants of the probability density function of a noise. In the simplest case of a
linear symplectic space the quadratic in momenta Hamiltonian corresponds to a Gaussian
noise.

To avoid misunderstanding we note in advance that the result of stochastic deformation of a classical model, that we call stochastic mechanics, is not Nelson's stochastic mechanics \cite{Nel} developed to give a stochastic interpretation of quantum mechanics. As a practical matter, our approach can be regarded as a
gauged version of the operator approach to stochastic mechanics expounded in \cite{GLSM,Nam,Par,Ris,ZJ,MatGlas}. Such a formulation offers the advantage that one can apply the developed methods of quantum mechanics and field theory to stochastic
mechanics almost without any changing. In this context it is sufficient to mention, for
instance, the models with gauge symmetries. The methods of quantization of systems with
constraints (see for the introduction \cite{Dirac,GitTyut,HeTe}) adapted to stochastic
deformation give the technique of generation of stochastic models respecting not merely the global symmetries of the initial classical system but the gauge symmetries as well. In
other words, these methods allow us to introduce the noise to physical
degrees of freedom that are not expressed in an explicit form. We shall consider several
models with gauge symmetries in this article.

The paper is organized as follows. We start our investigation with a formulation of
general rules of stochastic deformation by the example of a linear symplectic space
(Section \ref{stq rules}). In developing this scheme we follow the central proposal of deformation quantization and try to devise the procedure in a more algebraic way. We do not deal with the linear space of states and its dual but with the algebra of operators on them which are recognized as stochastic observables. A state of a stochastic system is specified by an analog of the density operator. For pure states it is useful to realize the algebra of observables in some linear space, at that we construct this space by the action of creation operators on the vacuum state. This standpoint allows us to calculate averages and make some proofs by means of the basic relation of the Heisenberg-Weyl algebra only. Generally speaking, the realization of some operators arising on intermediate steps of calculations can lead to divergent integrals or series that cancel each other in the final result. Then the algebraic approach can be considered as some regularization or prescription to handle these singular integrals. Besides, on this algebraic way we can establish stochastic mechanics for systems with nonlinear phase space such as symplectic or even non-regular Poisson manifolds. Explicit constructions of the deformed product (star-product) of the algebra of smooth functions on such manifolds are given in \cite{Fed,Kont}.

In conclusion of Section \ref{stq rules} we relate the mechanics obtained from stochastic deformation and ordinarily formulated stochastic mechanics using a path-integral representation of the transition probability and the Langevin equations associated with it. There we present the formal relation only. The existence problems are left beyond  the scope of the paper. The very method of stochastic deformation implies that the partial differential equations of the Fokker-Planck type are primary for this framework, while the stochastic equations of the Langevin type are only used to give a more lucid physical interpretation to the obtained equations. At the same time the form of the Hamiltonian generating an evolution is related to the probability density function of the noise. Thereby the information about the probability distribution law for the noise enters to the stochastic deformation procedure.

In Section \ref{examples} we consider several significant models revealing the key
features of the developed formalism. All of them are divided into three categories. The
first class exposed in Section \ref{nonrel part} is constituted by the models of a
nonrelativistic particle under the influence of a random force. It is a classical subject
of stochastic mechanics and we consider these models to include them into a general
scheme and gain an experience required in understanding of subsequent sections. Namely,
in this section we study the stochastic deformation of the models of a nonrelativistic
particle coupled to the electromagnetic and gravitational fields as well as the
stochastic deformation of the model leading to the Klein-Kramers equation. Under
stochastic deformation the first two models result in the stochastic systems described by the Fokker-Planck equations with trivial and nontrivial diffusion tensors respectively,
the diffusion tensor being the inverse metric \cite{Grah}.

In Section \ref{rel part sec} we extend our analysis to relativistic models with
constraints. We investigate three stochastic models of a relativistic particle affected
by a random force. The first model is the stochastically deformed model of a relativistic particle interacting with the electromagnetic field. We show that such a model is
described by a relativistic equation which generalizes the Fokker-Planck equation in the
same sense as the Klein-Gordon equation generalizes the Schr\"{o}dinger equation. For a
constant systematic force acting on the particle the obtained equation is in one-to-one
correspondence with the relativistic diffusion equation \cite{JosPre}.
Making use of the descent method \cite{Cour,Vlad} we derive a path-integral
representation of the transition probability, whence it turns out that such a
relativistic Fokker-Planck equation corresponds to a relativistic particle under the
influence of a non-Gaussian noise. The second model examined in this section is a
relativistic generalization of the model leading under stochastic deformation to the
Klein-Kramers equation. The respective relativistic equation proves to describe a
relativistic diffusion studied in \cite{DebMalRiv,DunHan,Fa,AngFra,ChAcKr,CCPDTH}. Here we
also obtain a path-integral representation of the transition probability both in the
gauge of a laboratory time and in the proper time gauge. The third model of this section
is related to stochastic mechanics of a relativistic charged particle with the radiation
reaction taken into account, i.e., we derive in this section a relativistic analog of the Fokker-Planck equation associated with the stochastic Lorentz-Dirac equation. It looks
very likely that the procedure developed in these sections is straightforwardly generalized
to stochastic reparameterization invariant relativistic ordinary differential equations
of an arbitrary order.

Section \ref{rel fields} is devoted to stochastic deformation of relativistic field
theories. We only touch the problem and consider free models of the scalar and
electromagnetic fields. Under the assumptions of causality and relativistic invariance we derive the propagators for these models. In the case of the electromagnetic field the
obtained propagator coincides with the well-known correlator of the electromagnetic
fields in a transparent medium \cite{LandLifstat}. Besides the last model illustrates how the BRST-quantization technique can be applied to stochastic mechanics.

Thus we see that the stochastic deformation method covers a considerable part of
stochastic physics and apparently any other stochastic model can be formulated in terms
of this unifying algebraic approach. The first model in Section \ref{rel part sec} shows that the non-Markovian processes can be obtained by stochastic deformation as well. The non-Markovian processes have to be described by the models with constraints. Inasmuch as we regard in this paper many topics of theoretical physics the reference list is, of course, incomplete. I tried to make references to the basic works known to me and to the works
that might be useful to the reader in understanding and further developing of the scheme
evolved here.

We use the following notation and conventions. The system of units is chosen so that the velocity of light $c=1$. Greek letters denote space-time indices and Latin indices
indicate the spatial components of tensors. Sometimes we shall use boldface characters to denote the spatial part of coordinates. Einstein's summation rule is assumed unless
otherwise stated. Usually overdots will denote a differentiation with respect to time. In the sections regarding nonrelativistic models $d$ is a dimension of the configuration
space with the coordinates $x$ and indices are risen and lowered by the metric tensor
$\eta_{\mu\nu}=diag(-1,1,\ldots,1)$, while for the relativistic models $d$ is a dimension of the space-time, $x$ is the space-time coordinates and the metric tensor
$\eta_{\mu\nu}$ reads as $diag(1,-1,\ldots,-1)$.


\section{The rules of stochastic deformation}\label{stq rules}

In this section we formulate the rules of stochastic deformation and establish the
relation between a stochastically deformed model and a certain stochastic
mechanics\footnote{The general scheme evolved in this paper was sketched in the preprint
\cite{hep}.}.

Consider a classical system on the linear symplectic manifold
$M$ with canonical coordinates $x^i$ and $p_j$:
\begin{equation}
    \{x^i,p_j\}=\de^i_j,\qquad i,j=\overline{1,d},
\end{equation}
where $d$ is a dimension of the configuration space and curly brackets denote the Poisson
brackets. The algebra of classical observables is the real commutative associative
algebra of smooth functions on $M$. The evolution is generated by
the observable $H(t,q,p)$ that is the Hamilton function.

Then we deform the algebra of classical observables in the manner
of deformation quantization but with an imaginary
deformation parameter (the Planck constant) such that
\begin{equation}\label{comm rel}
    [\hat x^i,\hat p_j]=\hbar\de^i_j,\qquad\hbar>0.
\end{equation}
Hereinafter we denote elements of the deformed algebra by hats
and imply the Weyl-Moyal star-product\footnote{For the introduction to different orderings, symbols and star-products see, e.g., \cite{BerezMSQ}.}
\begin{equation}
    \hat{f}\hat{g}=\sum\limits_{n=0}^\infty\frac1{n!}\left(\frac{\hbar}2\right)^n\omega^{a_1b_1}\ldots\omega^{a_nb_n}\partial_{a_1\ldots
    a_n}f(z)\partial_{b_1\ldots b_n}g(z),
\end{equation}
where $z\equiv(x,p)$, $a_n,b_n=\overline{1,2d}$, the functions $f(z)$ and $g(z)$
are the Weyl symbols of the corresponding elements of the deformed
algebra, $\omega^{ab}$ is the inverse to the symplectic $2$-form
$\omega_{ab}$. Recall that the Weyl-ordered operator for a
monomial in momenta symbol looks like
\begin{equation}\label{weyl ordering}
    V^{i_1\ldots i_n}(x)p_{i_1}\ldots
    p_{i_n}\;\rightarrow\;\frac1{2^n}\sum\limits_{k=0}^nC^k_n\hat{p}_{i_1}\ldots\hat{p}_{i_k}V^{i_1\ldots
    i_n}(\hat{x})\hat{p}_{i_{k+1}}\ldots\hat{p}_{i_n},
\end{equation}
where $C_n^k$ are the binomial coefficients.
The physical meaning of the constant $\hbar$ in formula
\eqref{comm rel} will be elucidated below. Roughly, $\hbar$
characterizes a variance of stochastic forces acting on a
classical system.

Let us given a linear functional $\Sp$ on the deformed algebra
mapping to real numbers and vanishing on commutators, viz.
\begin{equation}
    \Sp(\hat{f}\hat{g})=\Sp(\hat{g}\hat{f}),\quad\forall\hat{f},\hat{g},
\end{equation}
which we shall call the trace. An explicit formula for the trace of
the element $\hat{f}$ has the form
\begin{equation}
    \Sp\hat f=\int{\frac{d^dxd^dp}{(2\pi\hbar)^d}f(x,ip)}\;\Rightarrow\;\Sp(\hat f\hat
    g)=\int{\frac{d^dxd^dp}{(2\pi\hbar)^d}f(x,ip)g(x,ip)}.
\end{equation}
Then we define a complete set of elements $\{\hat{\rho}_\la\}$ of
the deformed algebra by the properties
\begin{equation}
\begin{aligned}
    &1.&\quad &\sum\limits_\la\hat\rho_\la=\hat{1},\\
    &2.&\quad &\Sp\hat\rho_\la=1,\\
    &3.&\quad &\hat\rho_\la\hat\rho_{\la'}=\de_{\la\la'}\hat\rho_\la.
\end{aligned}
\end{equation}

Consider a class of the complete sets related to each other by
similarity transformations, i.e., two complete sets
$\{\hat\rho_\la\}$ and $\{\hat\s_\mu\}$ are in the same class iff
there exists an invertible element $\hat{U}$ in the algebra spanned on the generators
$\hat{x}^i$ and $\hat{p}_j$ such that
\begin{equation}
    \hat{\s}_{f(\la)}=\hat{U}^{-1}\hat{\rho}_\la\hat{U},
\end{equation}
where $f$ is a bijection and $\hat{U}$ does not depend on $\la$.
We choose the class which contains the complete set
$\{\hat\rho_x\}$ corresponding to the $x$-representation
\begin{equation}\label{x representation}
    \hat\rho_x=\de^d(\hat{x}^i-x^i)=\int{\frac{d^dp}{(2\pi\hbar)^d}\exp\left[{\frac{i}\hbar
p_i(\hat{x}^i-x^i)}\right]}.
\end{equation}
This class also contains the complete sets associated with any
Lagrangian section of the symplectic space $M$ obtained from the
coordinate Lagrangian section by a linear symplectic
transformation. The element $\hat U$ realizing a similarity
transformation is a solution of the equation
\begin{equation}
    \hbar\dot{\hat{U}}(t)=\frac12\hat{z}^a\omega_{ab}A^b_c\hat{z}^c\hat{U}(t),\qquad\hat{U}(0)=\hat{1},
\end{equation}
for appropriate $t$, where the matrix $A^a_b$ belongs to the Lie
algebra of the symplectic group.

We say that the element $\hat{T}$ of the deformed algebra is a
stochastic observable iff there exists a complete set
$\{\hat{t}_\la\}$ from the chosen class such that
\begin{equation}
    \hat{T}=\sum\limits_{\la}T(\la)\hat{t}_\la,
\end{equation}
where $T(\la)$ is a certain real-valued function. In other words
the stochastic observables should be diagonalizable in the chosen
class.

The state of the stochastic system is characterized by the
observable $\hat{\rho}$ with a unit trace:
\begin{equation}
    \Sp\hat\rho=1.
\end{equation}
The pure state is specified by an additional idempotency
requirement
\begin{equation}
    \hat{\rho}^2=\hat\rho.
\end{equation}
The average of an observable $\hat{T}$ over a state $\hat{\rho}$
is defined as
\begin{equation}
    \lan\hat T\ran:=\Sp(\hat\rho\hat T).
\end{equation}
In particular, the average over the state $\hat\rho$ of a certain
pure state $\hat\rho_\la$ gives the probability to find a stochastic system in
this pure state.

The dynamics of a stochastic system in the state $\hat{\rho}$ are
generated by the Weyl-ordered operator $\hat{H}$ corresponding to
the Hamilton function $H(t,x,p)$ and described by the von Neumann
equation
\begin{equation}\label{neumann eqs}
    \hbar\dot{\hat\rho}=[\hat
    H,\hat\rho],
\end{equation}
whence the ``Euclidean'' Heisenberg equation for averages follows
\begin{equation}\label{heis eqs}
    \hbar\frac{d}{dt}\lan\hat T\ran=\lan\hbar\partial_t\hat T+[\hat T,\hat
    H]\ran.
\end{equation}
The evolution defined in this way maps observables into
observables. Besides it follows from \eqref{neumann eqs} that the
evolution of a probability density function to find a system in
the pure state $\hat{\rho}_\la$ obeys the equation
\begin{equation}\label{Fokker-Planck eq}
    \hbar\frac{d}{dt}\lan\hat{\rho}_\la\ran_{\hat\rho}=\Sp(\hat\rho_\la[\hat
H,\hat\rho]),
\end{equation}
which is nothing but the Fokker-Planck equation. The conservation
of the total probability is a consequence of the trace property.

Now we are in position to touch the problem of an arbitrariness in
a definition of the star-product related to the ordering
prescription. This arbitrariness at least for a linear symplectic
space does not affect averages and their evolution, but as in
quantum mechanics when we convert one star-product to another
some observables becomes nondiagonalizable in a class of complete
sets corresponding to the new star-product. Anyway different
orderings for the same physical observable or the Hamiltonian just
result in different definitions of the quantities measured in stochastic
mechanics (correlators) in terms of the coefficients of an expansion
of the observable and the Hamiltonian in momenta.

Since the phase space of the given classical system is a linear
symplectic space it is useful to realize the deformed algebra by operators acting in
some linear subspace $V$ of the linear space of smooth functions on the configuration
space. We introduce the Dirac notations for elements from the linear space
$V$ and its dual:
\begin{equation}
    |\psi\ran\in V,\qquad\lan\vf|\in V^*,
\end{equation}
where $V^*$ is the linear space of linear functionals on $V$.
Operators acting in the space $V$ are naturally translated to
operators acting in the dual space $V^*$. The eigenvectors from
the space $V$ and $V^*$ of the operator $\hat{T}$ corresponding to
the eigenvalue $t$ we denote by
\begin{equation}
\begin{gathered}
    |T=t\ran,\;\text{or}\;|t\ran,\quad\text{and}\quad\lan T=t|,\;\text{or}\;\lan t|,\\
    \hat{T}|t\ran=t|t\ran,\qquad\lan t|\hat{T}=\lan t|t.
\end{gathered}
\end{equation}

For the given left and right vectors (vacua) $|x=0\ran$ and $\lan
x=0|$ we construct the eigenvectors of the operator $\hat{x}$
\begin{equation}
    |x=a\ran=e^{\frac1\hbar a^i\hat{p}_i}|x=0\ran,\qquad\lan x=a|=\lan
x=0|e^{-\frac1\hbar a^i\hat{p}_i}.
\end{equation}
Their inner product possesses the properties
\begin{equation}
    f(x)\lan x|x'\ran=f(x')\lan x|x'\ran,\quad\forall f\in V,\qquad\lan
x+\la|x'+\la\ran=\lan x|x'\ran,\quad\forall\la,
\end{equation}
and, consequently, is equal up to a constant factor to the
$\de$-function. On setting this constant to unity the element of
the complete set \eqref{x representation} can be written as
\begin{equation}\label{projector}
    \hat\rho_x=|x\ran\lan x|.
\end{equation}
In the same manner we can prove that an element of a complete set
associated with any other Lagrangian section has the from
\eqref{projector} with appropriate eigenvectors. The linear space $V$ is spanned on  the
vectors obtained from $|x\ran$ by an action of the various operators $\hat{U}$ realizing
the similarity transformation in the chosen class of complete sets.

Thus for the pure state
\begin{equation}\label{eq:pure state}
    \hat\rho=|\psi\ran\lan\vf|,\qquad\lan\vf|\psi\ran=1,
\end{equation}
the probability to find a system in the point $x$ at the time $t$
takes the form
\begin{equation}\label{eq:PDF}
    \rho(t,x)=\lan x|\psi\ran\lan\vf|x\ran.
\end{equation}
Now it is easy to obtain a formal solution to the Fokker-Planck
equation \eqref{Fokker-Planck eq}. Its fundamental solution or, in
physical terms, the transition probability can be represented in
two ways\footnote{Other representations are permissible if one
factorizes the $\de$-function, i.e.,
\[
    \de(x-y)=\vf(x,y)\psi(x,y),
\]
where $\vf(x,y)$ and $\psi(x,y)$ are not proportional to the
$\de$-function. This is possible, for example, in the
multidimensional case or by the use of exotic functions such as
the fractional powers of the $\de$-function (see, e.g., \cite{BudSam}). In
both cases one should take care about linearity of Eq.
\eqref{Fokker-Planck eq} with respect to $\rho(t,x)$.}
\begin{equation}\label{eq:trans prob}
    G(t',x';t,x):=\frac{\lan
x'|\hat{U}_{t',t}|x\ran\lan\vf|\hat{U}^{-1}_{t',t}|x'\ran}{\lan\vf|x\ran}=\frac{\lan
x'|\hat{U}_{t',t}|\psi\ran\lan x|\hat{U}^{-1}_{t',t}|x'\ran}{\lan x|\psi\ran},
\end{equation}
where $\hat{U}_{t',t}$ is the evolution operator obeying the
equations
\begin{equation}
    \hbar\partial_{t'}\hat{U}_{t',t}=\hat{H}(t')\hat{U}_{t',t},\qquad\hat{U}_{t,t}=\hat{1}.
\end{equation}
A convolution of \eqref{eq:trans prob} with \eqref{eq:PDF} gives
the probability density function at the moment $t'$. To provide
linearity of equations \eqref{Fokker-Planck eq} with respect to
$\rho(t,x)$ we have to claim that one of the kernels
\eqref{eq:trans prob} is independent of $\rho(t,x)$. Therefore
either $\lan\vf|x\ran$ or $\lan x|\psi\ran$ is independent of
$\rho(t,x)$.

Consequently, the density operator \eqref{eq:pure state} in the
$x$-representation looks like
\begin{equation}\label{eq:phase def}
    \lan x|\psi\ran\lan\vf|y\ran=e^{\frac1\hbar[S(t,y)-S(t,x)]}\rho(t,x),\qquad\lan
x|\psi\ran\lan\vf|y\ran=e^{\frac1\hbar[\tilde{S}(t,y)-\tilde{S}(t,x)]}\rho(t,y),
\end{equation}
for the first and second cases respectively. Here we introduce
$S(t,x)$ which is the analog of a quantum mechanical phase \cite{Zambr,BlGaOlk}. This function
can have discontinuities or even be complex but its gradient contributing to the
observable averages should be real and have removable discontinuities only. The
first representation in \eqref{eq:trans prob} for the fundamental
solution is a forward evolution operator twisted by
$e^{\frac1\hbar S(t,x)}$ and it takes the form of a forward
transition probability, while the second representation of the
fundamental solution has the form of a backward transition
probability for the Fokker-Planck equation.

These transition probabilities depend on the phase
having different definitions for the first and second cases
\eqref{eq:phase def}. For the given initial probability density
function one can choose the initial phase in such a way that the
forward transition probability will give the same probabilities
as the process generated by the backward transition probability
with the same initial probability density function. The phase is a new ``degree of freedom'' as against to classical mechanics and the above mentioned flexibility of the formalism is only concerned with its interpretation in terms of stochastic mechanical quantities. For the forward transition probability with the definition of phase given by the first formula in \eqref{eq:phase def} this interpretation takes a simplest form and will be discussed below. Henceforth we shall refer to the forward
transition probability as the transition probability.

\subsection{Stochastic deformation and Langevin equation}

 By
standard means (see, e.g., \cite{DemCh,BerezMSQ}) we can construct a path-integral representation
of the transition probability
\eqref{eq:trans prob}. Notice in advance that we shall not consider here the existence problems of fundamental solutions and their well-defined path-integral representations. The interested reader can consult the references above.

Using the relation between the kernel of
the Weyl-ordered operator $\hat{T}$ in the coordinate
representation and its symbol
\begin{equation}
    \lan
x'|\hat{T}|x\ran=\int{\frac{d^dp}{(2\pi\hbar)^d}T\left(\frac{x'+x}2,ip\right)e^{-\frac{i}\hbar
p_i(x'^i-x^i)}},
\end{equation}
we formally have
\begin{multline}\label{matrix element decomp}
    \lan\vf(t+dt)|x'\ran\lan x'|\hat U_{t+dt,t}|x\ran\frac1{\lan
    \vf(t)|x\ran}=\\
    \lan x'|\exp\left\{\frac{dt}\hbar\left[\hat{H}\left(t,\hat{x},\hat{p}+\nabla
S(t,\hat{x})\right)+\partial_tS(t,\hat{x}) \right] \right\}|x\ran=\\
    \int\frac{d^dp(t)}{(2\pi\hbar)^d}\exp{\left\{-\frac{i}\hbar\left[p_i(t)\dot{x}^i(t)+i\left(\bar{H}_W\left(t,\tilde{x}(t),ip(t)\right)+\partial_t
    S(t,\tilde{x}(t))\right)\right]dt\right\}},
\end{multline}
where $\bar{H}_W(t,x,p)$ is a Weyl-symbol of the Hamiltonian $\hat{H}$ in which the momenta operators $\hat{p}_j$ are replaced by $\hat{p}_j+\partial_j\hat
S$ and
\[
    x(t)=x,\qquad x(t+dt)=x',\qquad \dot{x}(t):=(x(t+dt)-x(t))/dt,\qquad\tilde{x}(t):=(x(t+dt)+x(t))/2.
\]
As long as the transition probability \eqref{eq:trans prob}
possesses the defining  property of the Markov process we can cut
the time interval $[t,t']$ into pieces for which formula
\eqref{matrix element decomp} makes sense and then integrate over
intermediate positions. As a result the fundamental solution to
the Fokker-Planck equation \eqref{Fokker-Planck eq} takes the form
\begin{multline}\label{trans prob func int Weyl}
    G(t',x';t,x)=\int
    \prod_{\tau\in(t,t')}d^dx(\tau)\prod_{\tau\in[t,t')}\frac{d^dp(\tau)}{(2\pi\hbar)^d}\times\\
    \exp\left\{-\frac{i}\hbar\int\limits_t^{t'-d\tau}d\tau\left[p_i(\tau)\dot{x}^i(\tau)+i\left(\bar
    H_W(\tau,\tilde{x}(\tau),ip(\tau))+\partial_\tau
    S(\tau,\tilde{x}(\tau))\right)\right]\right\}.
\end{multline}
Notice that the main contribution to the transition probability
\eqref{trans prob func int Weyl} is made by paths approximating a
classical trajectory. This becomes manifest if one makes the change
of variables $p_j\rightarrow p_j-\partial_j S$ and neglects
stochastic corrections.

On integrating \eqref{trans prob func int Weyl} over momenta we arrive
at the functional integral with the expression in the exponent taking the
form of the Stratonovich-type stochastic integral (see, e.g.
\cite{Ris}). It is a consequence of the use of Weyl-symbol for the evolution operator. If we had chosen the $pq$-symbol,
we would have obtained the Ito-type stochastic integral. Namely, in
that case the transition probability \eqref{trans prob func int
Weyl} becomes
\begin{multline}\label{trans prob func int}
    G(t',x';t,x)=\int
    \prod_{\tau\in(t,t')}d^dx(\tau)\prod_{\tau\in[t,t')}\frac{d^dp(\tau)}{(2\pi\hbar)^d}\times\\
    \exp\left\{-\frac{i}\hbar\int\limits_t^{t'-d\tau}d\tau\left[p_i(\tau)\dot{x}^i(\tau)+i\left(\bar
    H(\tau,x(\tau),ip(\tau))+\partial_\tau
    S(\tau,x(\tau))\right)\right]\right\},
\end{multline}
where $\bar{H}(t,x,p)$ is a $pq$-symbol of the Hamiltonian
operator with the momenta $\hat{p}_j+\partial_j\hat{S}$. Certainly, the transition
probabilities \eqref{trans prob func int
Weyl} and \eqref{trans prob func int} are equal to each other as
they are different symbols of the same evolution operator.

Let us now provide an interpretation of the above stochastic mechanics
in terms of the Langevin equation
\begin{equation}\label{eq:langevin}
    \dot{x}^i(\tau)-\nu^i(\tau)=0.
\end{equation}
To this end we insert the $\de$-function with the LHS of \eqref{eq:langevin} in its argument into the path-integral \eqref{trans prob
func int} and integrate it over $\nu^i(\tau)$. Then
\begin{equation}\label{eq:PDF for noise}
    F(\nu,\tau,x):=\int\frac{d^dp}{(2\pi\hbar)^d}e^{-\frac{i}\hbar d\tau
p_i\nu^i}\exp\left\{\frac{d\tau}{\hbar}\left[\bar
    H(\tau,x,ip)+\partial_\tau
    S(\tau,x)\right]\right\},
\end{equation}
can be interpreted as a probability density function of the noise
$\nu^i(\tau)$ at the time slice $\tau$, the Langevin equation
\eqref{eq:langevin} being understood in the Ito sense. The
function
\begin{equation}\label{charact func}
    \Phi(\la,\tau,x):=\int d^d\nu
e^{i\la_i\nu^i}F(\nu,\tau,x)=(d\tau)^{-d}\exp\left\{\frac{d\tau}{\hbar}\left[\bar
    H\left(\tau,x,i\la\frac{\hbar}{d\tau}\right)+\partial_\tau
    S(\tau,x)\right]\right\},
\end{equation}
is known as the characteristic function and the coefficients of
the Taylor series in $i\la$ of the expression in the exponent are
called the cumulants. For example, the second cumulant is
proportional to the inverse metric (inverse mass matrix) in the
Hamiltonian $\bar H(t,x,p)$ multiplied by the deformation
parameter $\hbar$. The second cumulant is equal to the
mean-squared deviation that is why we claimed that $\hbar$
characterizes the variance of stochastic forces acting on the
system.

Now it is easy to see that the Hamiltonian at most quadratic in
momenta leads to the Langevin equation with a noise whose
the probability density function is a product of delta and
Gaussian functions. The Hamiltonians depending on higher powers of
momenta correspond to a non-Gaussian noise. If the Hamiltonian is
analytic in momenta then in the classical limit,
$\hbar\rightarrow0$, only the first cumulant survives and the
transition probability \eqref{trans prob func int} becomes
Liouvillian. The zeroth cumulant is a mere normalization factor
and, as follows from the normalization condition on \eqref{trans
prob func int}, is equal to $-d\ln d\tau$.

The above interpretation in terms of the Langevin equation works well when the
probability density
\eqref{eq:PDF for noise} of noise is a positive and normalizable
function. If this function takes negative values then the transition
probability \eqref{trans prob func int} also possesses negative
values for some values of its arguments. However it does not
automatically imply that the considered stochastic mechanics is
unphysical. For example, this fact may signify that the
$\de$-localized probability density functions are not well defined
for such a system from the physical point of view. At the same
time there may exist a class of probability density functions for
which a convolution with the transition probability gives
reasonable results. We shall encounter with the problem of negative probabilities in considering a relativistic generalization of the Fokker-Planck equation in Section \ref{rFPE}.

\section{Examples}\label{examples}
\subsection{Nonrelativistic particle}\label{nonrel part}

In this subsection we consider the stochastic deformation of three nonrelativistic
models: a particle interacting with the electromagnetic field, the same model on a curved
background and the model leading to the Klein-Kramers equation. For these models we
obtain a path-integral representations of the transition probabilities and the associated
Langevin equations. Some simple applications of the developed scheme are also
demonstrated.

\subsubsection{Fokker-Planck equation}

According to general rules expounded in the previous section
the Weyl-ordered Hamiltonian for a nonrelativistic particle looks
like
\begin{equation}\label{hamiltonian}
    \hat{H}=\frac{(\hat{p}_i-\hat{A}_i)^2}{2m}+\hat{A}^0.
\end{equation}
It is convenient to realize the operators $\hat{x}^i$ and
$\hat{p}_j$ in the linear space $V$ of functions on the configuration
space in the following way
\begin{equation}
    \hat{x}^i=x^i,\qquad\hat{p}_i=-\hbar\partial_i.
\end{equation}
Then for the pure state of the form \eqref{eq:pure state} the von
Neumann equation \eqref{neumann eqs} reduces to the two ``Euclidean''
Schr\"{o}dinger equations
\begin{equation}\label{shrodinger eqs nonrel}
    \hbar\partial_t\psi(t,x)=\left[\frac{(\hat{p}_i-A_i)^2}{2m}-A_0\right]\psi(t,x),\qquad\hbar\partial_t\vf(t,x)=-\left[\frac{(\hat{p}_i+A_i)^2}{2m}-A_0\right]\vf(t,x),
\end{equation}
provided that the standard inner product is understood. For a correct stochastic interpretation the functions $\psi(t,x)$ and $\vf(t,x)$ have to be both positive (or both negative). The fields $A_\mu$ are the gauge fields, which we shall call the
electromagnetic fields. Their physical meaning is obvious from the
general considerations of the previous section. Namely, introducing the phase, $S:=\hbar\ln\vf$, we see from \eqref{charact func} that the quantity
\begin{equation}
    \partial_iS-A_i
\end{equation}
is proportional to the first cumulant of
the probability density function of noise \eqref{eq:PDF for noise} whereas the first cumulant is equal to the
expectation value of a random variable. We dwell on the interpretation of $A_\mu$ a bit
later.

The Schr\"{o}dinger equations \eqref{shrodinger eqs nonrel} are invariant under the
following gauge transformations
\begin{equation}\label{gauge trans}
    \psi(t,x)\rightarrow\psi(t,x) e^{-\xi(t,x)},\qquad\vf(t,x)\rightarrow\vf(t,x)
e^{\xi(t,x)},\qquad A_\mu(t,x)\rightarrow
    A_\mu(t,x)+\partial_\mu\xi(t,x).
\end{equation}
In particular, these transformations do not change the probability
density function. The conserved $4$-current corresponding to the
gauge transformations \eqref{gauge trans} is given by
\begin{equation}\label{eq:4 current}
    j^\mu=\left(\vf\psi,\frac1{2m}\left[\vf(\hat{p}^i-A^i)\psi-\psi(\hat{p}^i+A^i)\vf\right]\right).
\end{equation}
The system of equations \eqref{shrodinger eqs nonrel} is
Lagrangian with the Hamiltonian action of the form
\begin{equation}\label{eq:action shr eqs}
    S_H[\vf,\psi]=\int
    dt\lan\vf|\hbar
    \partial_t-\hat{H}|\psi\ran,
\end{equation}
that is the fields $\psi(t,x)$ and $\vf(t,x)$ are canonically
conjugate with respect to the Poisson bracket.

On substituting the definition of phase \eqref{eq:phase def}, the
system of evolutionary equations \eqref{shrodinger eqs nonrel} can
be rewritten in the equivalent form
\begin{equation}\label{FP and qHJ eqs}
    \partial_t\rho=-\partial^i\left[-\frac{\hbar}{2m}\partial_i\rho+\frac{\partial_iS-A_i}m\rho\right],\qquad\partial_tS-A_0+\frac{(\partial_i
    S-A_i)^2}{2m}=-\frac{\hbar}{2m}\partial^i(\partial_i
    S-A_i).
\end{equation}
The first equation is Fokker-Planck equation,
which is the divergence of the $4$-current \eqref{eq:4 current},
while the second equation can be referred to as the quantum
Hamilton-Jacobi equation \cite{BlGaOlk,LMSh}. Now it is evident that if
one neglects stochastic corrections then the initially
$\de$-shaped probability density function $\rho(t,x)$ keeps its
own form and propagates as a classical charged particle in the
electromagnetic field\footnote{Such an interpretation for the
Langevin equation with a non-conservative force was proposed in
\cite{LepMa}.} with particle's momentum
$\partial_iS(t,x)-A_i(t,x)$.

Taking into account the stochastic corrections we see that the
more the probability density function is localized, the higher is the
probability flow resisting localization. It can be perceived, for
example, from the stochastic analog of the quantum mechanical
uncertainty relation
\begin{equation}
    \lan(x^i)^2\ran\lan(p^{i}_{os})^2\ran\geq\frac{\hbar^2}4,
\end{equation}
where summation is not understood,
$p^i_{os}:=-\hbar\partial^i\ln\rho^{1/2}$ is the osmotic momentum.
This relation is easily deduced from the inequality
\begin{equation}
    \int d^dx\left[(\xi
    x^i-\hbar\partial_i)\rho^{1/2}\right]^2\geq0,\quad\forall\,\xi\in\mathbb{R},
\end{equation}
under the assumption that $\rho(x)$ tends to zero at spatial
infinity faster than $x^{-2}$.

To find the first stochastic correction to the classical equations of motion we use the
Heisenberg equations \eqref{heis eqs} for averages of the
operators $\hat{x}^i$ and $\hat{p}_j$
\begin{equation}
    m\frac{d}{dt}\lan
    x_i\ran=\lan\hat{p}_i-A_i\ran=\lan\partial_i
    S-A_i\ran,\qquad\frac{d}{dt}\lan\hat{p}_i\ran=-\lan\partial_iA^0\ran+\frac1{2m}\lan\partial_iA_j(\hat{p}_j-A_j)+(\hat{p}_j-A_j)\partial_iA_j\ran,
\end{equation}
whence
\begin{equation}\label{Newton eqs}
    m\frac{d^2}{dt^2}\lan x_i\ran=\lan
    E_i\ran+\frac1{m}\e_{ijk}\lan(\partial_j
    S-A_j)H_k\ran+\frac\hbar{2m}\e_{ijk}\lan\partial_jH_k\ran.
\end{equation}
In the case where $\rho(t,x)$ is sufficiently localized compared
to the characteristic scale of variations of the electromagnetic
fields the angle brackets can be carried through the
electromagnetic fields to obtain a closed system of evolutionary
equations on the average position. They are the Newton equations
with the stochastic correction.

Now we return to the interpretation of the gauge fields $A_\mu$.
Recall that the expectation value of noise in the Langevin
equation \eqref{eq:langevin} is called a systematic drift. In
our case the systematic drift is equal to
\begin{equation}\label{brownian eq1}
    f_i(t,x):=\partial_iS(t,x)-A_i(t,x),
\end{equation}
where we set $m=1$. Therefore to satisfy the second equation in
\eqref{FP and qHJ eqs} we have to take
\begin{equation}\label{brownian eq2}
    A_0-\partial_tS=\frac12\left(f^2+\hbar\partial_if^i\right).
\end{equation}
The system of equations \eqref{brownian eq1}, \eqref{brownian eq2}
with respect to $A_\mu(t,x)$ and $S(t,x)$ obviously admits a
solution for any systematic drift $f_i(t,x)$. The fields $A_\mu(t,x)$ and $S(t,x)$ are
not uniquely defined by these equations and the arbitrariness in their definition
is equivalent to the arbitrariness of a gauge. In particular, in
the ``unitary'' gauge $S(t,x)=0$ the gauge fields $A_i$ are the
components of the systematic drift with an opposite sign.

The Newton equations \eqref{Newton eqs} for the average position of the
particle in the representation \eqref{brownian eq1},
\eqref{brownian eq2} become
\begin{equation}\label{Newton eqs new repr}
    \frac{d}{dt}\lan
    x^i\ran=\lan f^i\ran,\qquad
    \frac{d^2}{dt^2}\lan x^i\ran=\lan
    (\partial_t+f^j\partial_j)f^i\ran+\frac\hbar2\lan\Delta f^i\ran.
\end{equation}
For example, if $f^i(t,x)$ is the velocity field of an
incompressible viscous fluid with the specific pressure $p$ and
kinematic viscosity $\nu$ (see, e.g., \cite{LandLifsh_hyd}) then
the second equation in \eqref{Newton eqs new repr} is replaced by
\begin{equation}
    \frac{d^2}{dt^2}\lan x_i\ran=-\lan
    \partial_ip\ran+(\nu+\hbar/2)\lan\Delta f_i\ran,
\end{equation}
i.e., the acceleration of mean position of the Brownian particle
is the same as for the particle which is not influenced by
stochastic forces but entrained by a fluid with a higher
viscosity.

To gain a better physical insight into the stochastically deformed
model of a nonrelativistic particle we construct the functional
integral representation \eqref{trans prob func int Weyl} of the
transition probability. The Weyl-symbol of the Hamiltonian with
the momenta $\hat{p}_j+\partial_jS$ arising in formula
\eqref{trans prob func int} is
\begin{equation}
    \bar{H}(t,x,ip)=\frac1{2m}\left[-p^2+2ip^i(\partial_iS-A_i)+(\partial_iS-A_i)^2
    \right]+A^0.
\end{equation}
Substituting this expression into \eqref{trans prob func int Weyl}
and integrating over momenta we arrive at
\begin{multline}\label{trans prob func int nonrel}
    G(t',x';t,x)=\int
    \left(\frac{m}{2\pi\hbar
d\tau}\right)^{d/2}\prod_{\tau\in(t,t')}\left(\frac{m}{2\pi\hbar
d\tau}\right)^{d/2}d^dx(\tau)\times\\
    \exp\left\{-\frac1\hbar\int\limits_t^{t'-d\tau}d\tau\left[\frac{m}2\dot{x}^2+(A_i-\partial_iS)\dot{x}^i-(A^0+\partial_\tau
S)\right]\right\},
\end{multline}
where the functions $A_\mu(t,x)$ and $S(t,x)$ obey the quantum
Hamilton-Jacobi equation \eqref{FP and qHJ eqs} and are taken at
the point $(t,x)=(\tau,\tilde{x}(\tau))$. Since the expression in
the exponent is the classical action modulo boundary terms the
main contribution to the transition probability is made by the
paths approximating a classical trajectory that is in agreement
with our general considerations. In the representation
\eqref{brownian eq1}, \eqref{brownian eq2} the transition
probability \eqref{trans prob func int nonrel} reduces to the well-known result
\begin{multline}\label{trans prob func int nonrel repr}
    G(t',x';t,x)=\int
    \frac{1}{(2\pi\hbar d\tau)^{d/2}}\prod_{\tau\in(t,t')}\frac{d^dx(\tau)}{(2\pi\hbar
d\tau)^{d/2}}\times\\
    \exp\left\{-\frac1{2\hbar}\int\limits_t^{t'-d\tau}d\tau\left[(\dot{x}(\tau)-f(\tau,\tilde{x}(\tau)))^2+\hbar\partial_if^i(\tau,\tilde{x}(\tau))\right]\right\}.
\end{multline}
So we have obtained the transition probability by the use of the Weyl-symbol of the
evolution operator. The transition probability in terms of the $pq$-symbol \eqref{trans
prob func int} of the evolution operator is constructed along the same lines and looks
like \eqref{trans prob func int nonrel repr} without the stochastic correction to the
action and the midpoint prescription.

In conclusion of this section we briefly discuss how
different generalizations of the Fokker-Planck equation of the
form \eqref{FP and qHJ eqs} can be constructed in the developed
framework.

\subsubsection{Fokker-Planck equation with nontrivial diffusion tensor}

At first we consider stochastic deformation of the model
\eqref{hamiltonian} on a curved space background with the inverse
metric $g^{ij}$. From the general considerations we know that the
inverse metric is proportional to the covariance (the second
cumulant) of noise in the Langevin equation
\eqref{eq:langevin}, while the gauge fields $A_i$ are related to expectation values of the noise.
Besides, to provide a convergence of the integral \eqref{eq:PDF for
noise} over momenta we have to require that the eigenvalues of
the matrix $g^{ij}$ are nonnegative.

In constructing the Hamiltonian by its symbol we follow the simplest
prescription \cite{Und} based on the use of the exponential map
from the tangent bundle to the configuration space generated by
the Levi-Civita connection (for more sophisticated methods see,
e.g., \cite{BorNeuWal,GLS}). In other words we recognize the momenta $\hat{p}_i$ as
the derivatives at the origin of the Riemann normal coordinates
and the vectors of the linear space $V$ as the scalar functions on
the manifold. Neglecting for a while the gauge fields $A_\mu$ we
have for the Hamiltonian \eqref{hamiltonian} in the normal
coordinates
\begin{equation}
    \hat{H}_0=\frac1{8m}(\hat{p}_i\hat{p}_jg^{ij}+2\hat{p}_ig^{ij}\hat{p}_j+g^{ij}\hat{p}_i\hat{p}_j).
\end{equation}
Making use of relations between the derivatives of the metric
taken at the origin of the normal coordinates and the Riemannian
tensor \cite{Petrov,HarHaw} we arrive at\footnote{We use the sign
convention for the curvature as in \cite{Petrov,HarHaw}.}
\begin{equation}\label{eq:hamiltonian curv}
    \hat{H}_0=\frac{\hbar^2}{2m}\left(\nabla^2-\frac{R}{12}\right),
\end{equation}
where $R$ is the scalar curvature and $\nabla_i$ is the
Levi-Civita connection. Notice that the covariant $pq$-ordering
leads to the coefficient $1/3$ at the curvature as in
\cite{HarHaw}, the covariant $qp$-ordering gives rise to the vanishing
coefficient (minimal coupling) and
\[
    \hat{H}_0=\frac1{4m}(\hat{p}_i\hat{p}_jg^{ij}+g^{ij}\hat{p}_i\hat{p}_j)
\]
corresponds to \eqref{eq:hamiltonian curv} with the coefficient
$1/6$ at the curvature (conformal coupling) \cite{DeW}. All of
these prescriptions result effectively in a variation of the
coefficient at the scalar curvature.

Thus the Hamiltonian for a particle interacting with the gauge
fields $A_\mu$ on a curved space background looks like
\begin{equation}
    \hat{H}=\frac{1}{2m}(\hbar\nabla_i+A_i)^2-\frac{\hbar^2R}{24m}+A^0.
\end{equation}
Substituting the phase definition \eqref{eq:phase def} into the
Schr\"{o}dinger equations following from the action
\eqref{eq:action shr eqs} we obtain \cite{Grah}
\begin{equation}\label{eq:FP covariant}
\begin{gathered}
    g^{-1/2}\partial_t(g^{1/2}\rho)=-\nabla^i\left[-\frac{\hbar}{2m}\partial_i\rho+\frac{\partial_iS-A_i}m\rho\right],\\
    \partial_tS-A_0+\frac{(\partial_i
    S-A_i)^2}{2m}=-\hbar\partial_t\ln g^{1/2}-\frac{\hbar}{2m}\nabla^i(\partial_i
    S-A_i)+\frac{\hbar^2R}{24m},
\end{gathered}
\end{equation}
where $\rho$ and $S$ are assumed to be the
scalar functions and $g:=\det{g_{ij}}$. Since the Fokker-Planck equation is independent
of the scalar curvature term the averages of observables depending
on $x$ only and their evolution do not depend on the concrete
covariant ordering scheme.

Rewriting equations \eqref{eq:FP covariant} in terms of the
density $g^{1/2}\rho$ one can see that the inverse metric is
proportional to the diffusion matrix. Besides from the Heisenberg
equations we find a systematic drift
\begin{equation}
    m\frac{d}{dt}\lan x^i\ran=\lan
g^{ij}(\partial_jS-A_j)+\frac\hbar2g^{-1/2}\partial_j(g^{1/2}g^{ij})\ran.
\end{equation}
Despite the second term looks noncovariant it behaves under
general coordinate transformations like the first term under the assumption that
$\rho$ tends to zero at spatial infinity. The transition
probability \eqref{trans prob func int} takes the form (cf.
\cite{Grah,Weiss})
\begin{multline}
    G(t',x';t,x)=\int
    \prod_{\tau\in(t,t')}d^dx(\tau)\prod_{\tau\in[t,t')}\frac{d^dp(\tau)}{(2\pi\hbar)^d}\times\\
    \exp\left\{-\frac{i}\hbar\int\limits_t^{t'-d\tau}d\tau\left[p_i\dot{x}^i+\frac{i}m\left(-\frac{g^{ij}p_ip_j}{2}+ip_i\left(g^{ij}(\partial_jS-A_j)+\frac\hbar2g^{-1/2}\partial_j(g^{1/2}g^{ij})\right)\right)\right]\right\}.
\end{multline}
It is covariant under general coordinate transformations.
To prove this by making a change of variables one should take care
about the so-called extraterms (see, e.g., \cite{Prokh}) which
cancel derivatives of the Jacobian matrices resulting from the
noncovariant expression in the exponent.

\subsubsection{Klein-Kramers equation}

Now we turn to another generalization of the Fokker-Planck
equation \eqref{FP and qHJ eqs}, namely, to the Klein-Kramers
equation. This equation arises in studying systems of second order
stochastic differential equations. Therefore, firstly, introducing
additional variables we reduce such a system to the system of the
first order equations of the form \eqref{eq:langevin} and then
apply the developed formalism.

Let us consider a system of stochastic equations
\begin{equation}\label{eq:langevin KK}
    \dot{x}^i=\ups^i,\qquad\dot{\ups}^i=f^i(t,x,\ups)+\nu^i,
\end{equation}
where $\nu^i$ is a Gaussian white noise. In that case the points
of the configuration space are the pairs $(x,\ups)$. We denote by
$(p,\pi)$ canonically
conjugate variables to $(x,\ups)$ in the phase space
\begin{equation}
    \{x^i,p_j\}=\de^i_j,\qquad\{\ups^i,\pi_j\}=\de^i_j.
\end{equation}
From the representation \eqref{trans prob func int} it is not
difficult to see that this stochastic system is described
by the Hamiltonian
\begin{equation}\label{eq:ham KK}
    \hat{H}=\frac{(\hat{\pi}_i-\hat{A}_i)^2}2+\hat{\ups}^i\hat{p}_i+\hat{A}^0,
\end{equation}
where $A_\mu(t,x,\ups)$ are the gauge fields. The Hamiltonian
\eqref{eq:ham KK} is merely a general expression\footnote{A
specific form of the coefficients at $p_i$ is fixed by the classical
equations of motion for $x^i(t)$ which should coincide with
\eqref{eq:langevin KK}.} at most quadratic in momenta $\pi_i$ and
linear in momenta $p_i$. By introducing the phase $S(t,x,\ups)$ the
Schr\"{o}dinger equations corresponding to the Hamiltonian
\eqref{eq:ham KK} can be cast into the form
\begin{equation}\label{eq:KK qHJKK}
\begin{gathered}
    \partial_t\rho=-\partial^i_\ups\left[-\frac{\hbar}2\partial_i^\ups\rho+(\partial_i^\ups
S-A_i)\rho\right]-\partial^i_x(\ups_i\rho),\\
    \partial_tS-A_0+\frac{(\partial^\ups_iS-A_i)^2}2+\ups^i\partial_i^xS=-\frac\hbar2\partial^i_\ups(\partial_i^\ups
S-A_i).
\end{gathered}
\end{equation}
The first equation is the Klein-Kramers equation, while the second
equation is an analog of the Hamilton-Jacobi equation or, from
an utilitarian viewpoint, is the definition of $A_0$. These
equations are Lagrangian with the Hamiltonian action of the form
\eqref{eq:action shr eqs}. Besides they are invariant with respect
to the gauge transformations
\begin{multline}
    S(t,x,\ups)\rightarrow S(t,x,\ups)+\xi(t,x,\ups),\qquad
    A_i(t,x,\ups)\rightarrow
    A_i(t,x,\ups)+\partial_i^\ups\xi(t,x,\ups),\\
    A_0(t,x,\ups)\rightarrow
A_0(t,x,\ups)+\dot\xi(t,x,\ups)+\ups^i\partial_i^x\xi(t,x,\ups).
\end{multline}
The covariant derivatives respecting these gauge transformations
read as follows
\begin{equation}
    \hat{\mathcal{P}}_0=\hat{p}_0+\ups^i\hat{p}_i-A_0,\qquad\hat{\mathcal{P}}_i=\hat{\pi}_i-A_i.
\end{equation}
As far as an interpretation of the gauge fields $A_i$ is concerned
imposing the ``unitary'' gauge $S=0$ we see that these fields are
equal to the components of the systematic force $f_i$ with an
opposite sign. A path-integral representation of the transition
probability \eqref{trans prob func int} reads as
\begin{multline}
    G(t',x';t,x)=\int
    \prod_{\tau\in(t,t')}\left(d^dx(\tau)d^d\ups(\tau)\right)\prod_{\tau\in[t,t')}\frac{d^dp(\tau)d^d\pi(\tau)}{(2\pi\hbar)^{2d}}\times\\
    \exp\left\{-\frac{i}\hbar\int\limits_t^{t'-d\tau}d\tau\left[p_i\dot{x}^i+\pi_i\dot{\ups}^i+i\left(-\frac{\pi^2}{2}+i\pi^i(\partial^\ups_iS-A_i)+ip_i\ups^i\right)\right]\right\}.
\end{multline}

For the Hamiltonians at most quadratic in variables, i.e., for the
linear systematic force $f^i$ \cite{UlhOrn}, an evolution of the
stochastic system can be easily found from the Heisenberg
equations. Here we consider the simplest case
\begin{equation}\label{eq:syst force fric}
    f^i=-\ga\ups^i,
\end{equation}
just to illustrate how the formalism works. For the force
\eqref{eq:syst force fric} we can choose
\begin{equation}
    S=-\ga\frac{\ups^2}2,\quad A_i=0\;\Rightarrow\;A_0=\ga^2\frac{\ups^2}2-\frac\hbar2\ga
d,
\end{equation}
whence in the Heisenberg representation it follows from the definition of
the phase $S$ that
\begin{equation}\label{eq:left st fric}
    \lan\vf|\hat{\pi}_i(0)=-\ga\lan\vf|\hat{\ups}_i(0),\qquad\lan\vf|\hat{p}_i(0)=0.
\end{equation}
A general solution of the Heisenberg equations looks like
\begin{equation}\label{eq:heis eqs fric}
\begin{gathered}
    \hat{p}(t)=\hat{p},\qquad
\hat{x}(t)=\hat{x}+\ga^{-2}(\hat{p}t-\hat{\pi})-\ga^{-1}(\hat{\ups}-\ga^{-2}\hat{p})\sinh\ga
t+\ga^{-2}\hat{\pi}\cosh\ga t,\\
    \hat{\pi}(t)=\hat{\pi}\cosh\ga t+(\ga\hat{\ups}-\ga^{-1}\hat{p})\sinh\ga t,
\qquad\hat{\ups}(t)=\ga^{-2}\hat{p}+(\hat{\ups}-\ga^{-2}\hat{p})\cosh\ga
t+\ga^{-1}\hat{\pi}\sinh\ga t,
\end{gathered}
\end{equation}
where all the operators at the RHS of these equations are taken at
$t=0$. Making use of equations \eqref{eq:left st fric},
\eqref{eq:heis eqs fric} and the commutation relations one can
obtain an evolution of the average of any observable. For
instance,
\begin{equation}
    \frac{m}2\lan\ups^2(t)\ran=\frac{m\hbar}{4\ga}d+\left(\frac{m}2\lan\ups^2\ran-\frac{m\hbar}{4\ga}d\right)e^{-2\ga
t}.
\end{equation}
Then assuming that $\ga>0$ and the equipartition law is fulfilled
we determine the deformation parameter in equations \eqref{eq:KK
qHJKK}
\begin{equation}
    \hbar=\frac{2\ga kT}{m}.
\end{equation}
It is interesting to note that the trace of the equipartition law
in the Fokker-Planck equation \eqref{FP and qHJ eqs} with the
Hamiltonian \eqref{hamiltonian} is the relation
\begin{equation}\label{equipart law}
    \lim_{dt\rightarrow0}T\{\frac{m\dot{\hat{x}}^2(t)}2dt\}=\frac{m}{2\hbar_{FP}}[\hat{x}_i,[\hat{x}^i,\hat{H}]]=\frac{\hbar_{FP}}2d,
\end{equation}
where $\hat{x}(t)$ are the position operators in the Heisenberg
representation, $T$ means the chronological ordering and $\hbar_{FP}$ is the deformation
parameter in the Fokker-Planck equation. To
reproduce the equipartition law one should formally put
$dt=2\ga^{-1}$.


\subsection{Relativistic particle}\label{rel part sec}

In this subsection we consider three relativistic models which
under stochastic deformation give rise to relativistic
generalizations of the Fokker-Planck and Klein-Kramers equations,
and to the equation describing a massive charged particle influenced by external
systematic and stochastic forces with
the radiation reaction taking into account. In other words in the last
case a dissipation is described by the Lorentz-Dirac force
\cite{Lor,Dir}.

\subsubsection{Relativistic Fokker-Planck equation}\label{rFPE}

In the previous subsection we saw that the stochastically deformed
model of a nonrelativistic particle results in the Fokker-Planck
equation. Therefore it is reasonable to expect that stochastic
deformation of the model of a relativistic particle gives some
relativistic generalization of the Fokker-Planck equation.

The Hamiltonian action of an interacting relativistic particle has
the form
\begin{equation}\label{rel part}
    S_H[x,p,\la]=\int d\tau[p_\mu\dot{x}^\mu+\frac{\la}2(\mathcal{P}^2-m^2)],
\end{equation}
where $\mathcal{P}_\mu:=p_\mu-A_\mu$ and $A_\mu$ are the gauge fields.
The dynamics of the model \eqref{rel part} are governed by one first class
constraint. In the proper time parameterization
$\la=m^{-1}$ the evolution of the position of the particle obeys
the equation
\begin{equation}
    m\dot{x}^\mu=A^\mu-p^\mu.
\end{equation}
Thus by analogy with a nonrelativistic model one should expect
that the corresponding stochastic equations look in the unitary
gauge like
\begin{equation}\label{rel stoch eq 1}
    m\dot{x}^\mu=A^\mu+\ldots,
\end{equation}
where dots denote terms vanishing at $\hbar=0$, i.e., in the
classical limit $A^\mu$ plays the role of the $4$-momentum of the
particle. The problem that one encounters by naive introduction the noise in the RHS of Eq. \eqref{rel stoch eq 1} is to preserve the reparameterization invariance. In the proper time gauge, for example, we must guarantee the fulfillment of $\dot{x}^2=1$. The stochastic deformation procedure allows us to obtain the Fokker-Planck type equation associated with \eqref{rel stoch eq 1} respecting the gauge invariance.

We shall deform the model
\eqref{rel part} in the gauge of a laboratory time
\begin{equation}\label{eq:temp gauge}
    \tau=x^0.
\end{equation}
A thorough description of quantization of a relativistic particle
in the gauge \eqref{eq:temp gauge} is presented in \cite{GavGit} and
we just trace some basic steps of this procedure which are
necessary for us. Firstly, we solve the mass-shell constraint with
respect to the energy\footnote{Here we work in the particle
sector only. Antiparticle's states in the sense of \cite{GavGit} are not included.}
\begin{equation}\label{eq:constraint}
    \mathcal{P}_0-\sqrt{m^2+(\mathcal{P}_i)^2}=0.
\end{equation}
Then we naturally realize the Heisenberg-Weyl algebra of operators
in the linear space $V$ of two-component real vectors
\begin{equation}
    \Psi=\left[%
\begin{array}{c}
  \psi'(x) \\
  \psi(x) \\
\end{array}%
\right]\in V,\qquad \hat{x}^\mu:=\left[%
\begin{array}{cc}
  x^\mu & 0 \\
  0 & x^\mu \\
\end{array}%
\right],\qquad\hat{p}_\mu:=\left[%
\begin{array}{cc}
  -\hbar\partial_\mu & 0 \\
  0 & -\hbar\partial_\mu \\
\end{array}%
\right].
\end{equation}
Define the dual linear space $V^*$ as the space of real linear
functionals acting on elements of $V$ by the rule
\begin{equation}\label{eq:inner prod rel}
    \lan\Phi|\Psi\ran:=-\int{d\mathbf{x}}\Phi^+\left[%
\begin{array}{cc}
  0 & 1 \\
  1 & 0 \\
\end{array}%
\right]\Psi=-\int{d\mathbf{x}[\vf\psi'+\vf'\psi]},
\end{equation}
where
\begin{equation}
    \Phi=\left[%
\begin{array}{c}
  \vf'(x) \\
  \vf(x) \\
\end{array}%
\right].
\end{equation}
The pure states are as usual the projectors of the
form\footnote{As in Section \ref{stq rules} one can construct
elements of the linear spaces $V$ and $V^*$ formally acting by the
operators $\hat{p}_\mu$ on the doubly degenerate vacuum of the
operators $\hat{x}^\mu$.} \eqref{eq:pure state}.

The above construction allows us to realize the constraint
\eqref{eq:constraint} without going to pseudo-differential
operators:
\begin{equation}
    \hat{T}:=\hat{\mathcal{P}}_0-\left[%
\begin{array}{cc}
  0 & m^2+(A_i-\hbar\partial_i)^2 \\
  1 & 0 \\
\end{array}%
\right].
\end{equation}
The dynamics of a pure state of the stochastic system are governed
in the gauge \eqref{eq:temp gauge} by the two Schr\"{o}dinger
equations
\begin{equation}\label{eq:rel shrod eqs}
    \hat{T}|\Psi\ran=\lan\Phi|\hat{T}=0.
\end{equation}
In quantum mechanics they are simply the Klein-Gordon equation in
the first order formalism \cite{BagGit}. In our case these
equations are equivalent to
\begin{equation}\label{KG eqs}
    (\hat{\mathcal{P}}^2-m^2)\psi=0,\quad\psi'=\hat{\mathcal{P}}_0\psi,\qquad(\hat{\mathcal{P}}^{+2}-m^2)\vf=0,\quad\vf'=\hat{\mathcal{P}}^+_0\vf,
\end{equation}
where the cross denotes a formal conjugation. When the gauge fields
vanish these equations are the Klein-Gordon equations for
tachyons\footnote{For the interrelation between relativistic
random walk models and relativistic wave equations see, e.g.,
\cite{GJKSch,RanMug}.}. On the solutions of the
Schr\"{o}dinger equations \eqref{eq:rel shrod eqs} the inner
product \eqref{eq:inner prod rel} reduces to
\begin{equation}\label{eq:inner prod rel red}
    \lan\Phi|\Psi\ran=-\int{d\mathbf{x}[\vf\hat{\mathcal{P}}_0\psi+\psi\hat{\mathcal{P}}^+_0\vf]}.
\end{equation}
The probability density function corresponding to the state
$|\Phi\ran\lan\Psi|$ is proportional to the integrand of
\eqref{eq:inner prod rel red}.

The Heisenberg equations associated with the Schr\"{o}dinger
equations \eqref{eq:rel shrod eqs} allow to obtain the evolution
of an average of any physical observable and, in particular, to
establish a stochastic interpretation of the gauge fields $A_\mu$.
However we shall act in a different way to derive a direct
stochastic interpretation of Eqs. \eqref{KG eqs} in terms of the
Langevin equation of the form \eqref{eq:langevin}. To this aim we,
firstly, represent the fields $\vf(x)$ and $\psi(x)$ as
\begin{equation}
    \vf(x)=e^{\frac1\hbar S(x)},\qquad\psi(x)=\tilde\rho(x)e^{-\frac1\hbar S(x)},
\end{equation}
substitute them into Eqs. \eqref{KG eqs} and fix the unitary gauge
$S=0$. Then we have\footnote{Notice that for the constant fields $A_\mu$ Eq. \eqref{FP
and qHJ rel eqs} reduces to the well-known relativistic diffusion equation
\cite{JosPre}.}
\begin{equation}\label{FP and qHJ rel eqs}
    \partial^\mu\left[\frac\hbar2\partial_\mu\tilde\rho+A_\mu\tilde\rho\right]=0,\qquad
A^2-m^2=\hbar\partial^\mu A_\mu.
\end{equation}
The first equation is the conservation law of the probability
density current
\begin{equation}\label{eq:prob dens cur}
    j_\mu=\frac1m\left(A_\mu+\frac\hbar2\partial_\mu\right)\tilde\rho,
\end{equation}
while the second equation is the deformed mass-shell condition or
the mere definition of $A^0(x)$. For given $A_i(x)$ we take the
solution of the second equation in \eqref{FP and qHJ rel eqs}
which is regular in $\hbar$ and possesses the following classical
limit
\begin{equation}\label{eq:a zero}
    \left.A_0\right|_{\hbar=0}=\sqrt{m^2+(A_i)^2}.
\end{equation}
The solution defined in this way is unique. Consequently, if one
has $j_\mu(x)$ at the initial moment then one can find from Eqs.
\eqref{eq:prob dens cur} the function $\tilde\rho(x)$ and its first
derivative with respect to time at this initial moment and solve the Cauchy problem.

For the potential and stationary relativistic flow with a constant specific relativistic enthalpy \cite{LandLifsh_hyd} entraining the Brownian particle we can choose $A_i=\partial_i U(\spx)$, where $U$ is a Lorentz scalar. Then the stationary probability distribution function following from \eqref{eq:prob dens cur} looks like
\begin{equation}
    j^0=\sqrt{1+(A_i)^2/m^2}\exp(-2U/\hbar)/Z=\ga \exp(-2U/\hbar)/Z,\qquad Z=const,
\end{equation}
where $\ga$ is the Lorentz factor. The power of the $\ga$-factor is uniquely specified by the requirement that $j^0$ is the zeroth component of the $4$-vector.

It is easy to see from Eq. \eqref{eq:prob dens cur} that the derived equations \eqref{KG eqs}
lead to negative probabilities when the probability density
function changes rapidly on the spatial (time) scales less than or
comparable with the ``Compton wave length'' $\hbar m^{-1}$. This
is in a perfect analogy with the well-known property of the
Klein-Gordon equation in relativistic quantum mechanics
\cite{BjoDre}. Therefore Eqs. \eqref{KG eqs} have a correct
stochastic interpretation only for the fields with a
characteristic scale of variations much larger than $\hbar
m^{-1}$.

Despite the above remark we consider an evolution of the
probability density function localized in one point $x^i=y^i$ at
the initial moment $x^0=y^0$:
\begin{equation}\label{eq:PDF init rel}
    j^0(y^0,x^i)=\de^{d-1}(x^i-y^i).
\end{equation}
The solution of the Cauchy problem to the first equation in
\eqref{FP and qHJ rel eqs} is obtained via a convolution of the
retarded Green function with the initial distribution
\begin{equation}
    \left(1+\frac12a\hat{p}_0\right)\de^d(x-y),
\end{equation}
where the function $a(x)$ is the regular in $\hbar$ solution of
the equation
\begin{equation}
    \left(A_0+\frac\hbar2\partial_0\right)a(x)=1.
\end{equation}
Thus, formally, the transition probability is
\begin{equation}\label{eq:trans prob rel}
    G(x,y)=\left(A_0-\frac12\hat{p}_0\right)\frac{1}{\frac12\hat{p}^2-\hat{p}^\mu
A_\mu}\left(1+\frac12a\hat{p}_0\right)\de^d(x-y).
\end{equation}

The transition probability \eqref{eq:trans prob rel} does not
possess the defining property of the Markov process and it can not
be straightforwardly represented by a path-integral. To overcome
this difficulty we represent the transition probability
\eqref{eq:trans prob rel} as the sum of the transition
probabilities possessing the Markov property with respect to some new variable
(fictitious time). The stated procedure is an analog of what is called in relativistic
quantum mechanics the Fock's fifth parameter (proper time) formalism
\cite{Fock,Feyn}.

Under some technical assumptions on the transition probability
\eqref{eq:trans prob rel} the following equality for the operator
$\hat{G}$ with the kernel \eqref{eq:trans prob rel} holds
\begin{equation}\label{eq:hamiltonian rel}
    \hat{G}=\hbar^{-1}\int\limits_0^{\infty}{d\tau
e^{-\frac\tau\hbar\hat{H}}},\qquad\hat{H}:=\left(1+\frac12a\hat{p}_0\right)^{-1}\left(\frac12\hat{p}^2-\hat{p}^\mu
A_\mu\right)\left(A_0-\frac12\hat{p}_0\right)^{-1}.
\end{equation}
This formula is understood in the distributional sense. It is derived by an application
of the descent method (see, e.g., \cite{Cour,Vlad}) to the equation
\begin{equation}
    -\hbar\partial_\tau\psi(\tau,x)=\hat{H}\psi(\tau,x),\qquad\psi(0,x)=\de^d(x-y),
\end{equation}
and is valid when the integral \eqref{eq:hamiltonian rel} converges. It is not difficult
to prove the convergence of this integral in the case of the constant
fields $A_i(x)$. In a general case the integral \eqref{eq:hamiltonian rel} can be defined
at least perturbatively.

The evolution operator $\exp{[-\hbar^{-1}\tau\hat{H}]}$ respects the Markov property and
conserves the probability density function normalization. Its kernel in the
$x$-representation can be interpreted as the probability of particle's arrival to the
point $x$ of the space-time from the point $y$ with the value of the fictitious time
$\tau$. The sum over all such particles with different fictitious times gives the total
observable transition probability \eqref{eq:trans prob rel}.

Now we represent the kernel of the evolution operator $\exp{[-\hbar^{-1}\tau\hat{H}]}$ by
a path-integral along lines expounded in Section \ref{stq rules} and introduce the
corresponding Langevin equation \eqref{eq:langevin}. It seems impossible to find an
explicit simple expression for the probability density function of noise for
arbitrary fields $A_i(x)$ but some properties of this distribution are easily derived.

Firstly, in the classical limit $\hbar\rightarrow0$ the Hamiltonian \eqref{eq:hamiltonian
rel} reduces to
\begin{equation}
    \hat{H}=-\hat{p}^\mu A_\mu/A_0,
\end{equation}
where $A_0$ is specified by Eq. \eqref{eq:a zero}. Consequently, the probability density
function of noise reads
\begin{equation}
    F(\nu,\tau,x)\underset{\hbar\rightarrow0}{\longrightarrow}\de(\nu^0-1)\de^{d-1}\left(\nu^i-A^i/\sqrt{m^2+(A_i)^2}\right).
\end{equation}
We see that in the classical limit the fictitious time turns into the physical time $x^0$
of the lab frame. Besides the above formula gives the interpretation to the fields
$A_i(x)$, which confirms our expectations.

Secondly, in the nonrelativistic limit $A_0\approx m$ and $m^{-1}\|\hat{p}_0\|\ll1$,
where $\hat{p}_0$ is assumed to act on the probability density function,  the Hamiltonian
\eqref{eq:hamiltonian rel} is rewritten as
\begin{equation}
    \hat{H}=-\frac{(\hat{p}_i)^2}{2m}-\frac{\hat{p}^\mu A_\mu}{m},
\end{equation}
whence the standard Gaussian distribution for the noise follows. As in the classical
limit the fictitious time $\tau$ is the physical time $x^0$.

Thus Eqs. \eqref{KG eqs} generalize the Fokker-Planck equation
\eqref{FP and qHJ eqs} in the same sense as the Klein-Gordon
equation generalizes the Schr\"{o}dinger equation. In the classical limit these equations describe the same physical system and in the nonrelativistic limit Eqs. \eqref{KG eqs}
reduce to the Fokker-Planck equation \eqref{FP and qHJ eqs}. However it is worthy to note that the above interpretation in terms of the fictitious time has some weaknesses.
Namely, since we start from the probability density function \eqref{eq:PDF init rel}
localized in one point the obtained equations are ill-defined. This inevitably leads to
negative probabilities although they disappear in the classical and nonrelativistic
limits, when $\hbar m^{-1}$ tends to zero. Moreover the negative probabilities appear to
result in the probability density function of the noise acting on the particle arrived to the point $x$ with the fictitious time $\tau$ being not equal to zero in the superluminal region. But, of course, the sum over all the particles with different $\tau$ gives rise
to the transition probability with the support bounded by the light cone. A detailed
investigation of these peculiarities will be given elsewhere.

\subsubsection{Relativistic Klein-Kramers equation}

To formulate a relativistic generalization of the Klein-Kramers equation we start from
a classical relativistic system with the Hamiltonian action of the form (cf.
\eqref{eq:ham KK})
\begin{multline}\label{eq:action rKK}
    S_H[x,p,y,\pi,\la,\mu]=\int d\tau(p_\mu\dot{x}^\mu+\pi_\mu\dot{y}^\mu
    -\la T_1-\mu T_2),\\
    T_1:=\frac12g^{\mu\nu}(\pi_\mu+A_\mu)(\pi_\nu+A_\nu)+
    y^\mu p_\mu+\phi,\qquad
    T_2:=y^2-m^2,
\end{multline}
where dots denote the derivatives with respect to the parameter $\tau$, $m$ is a mass of
the particle, $A_\mu(x,y)$ and $\phi(x,y)$ are gauge fields, $g^{\mu\nu}(x,y)$ is a
symmetric Lorentz tensor orthogonal to $y_\mu$. In other words the dynamics of the system are governed by two first class constraints. The first constraint generates
reparameterizations of the world line of the particle, the second constraint being just the mass-shell condition. These two constraints are in the Abelian involution by virtue of the property of the tensor $g^{\mu\nu}$.

A part of the equations of motion in what we are interested in reads as follows
\begin{equation}
    \dot{x}^\mu=\la y^\mu,\qquad\dot{y}^\mu=\la g^{\mu\nu}(A_\nu+\pi_\nu),\qquad
y^2=m^2.
\end{equation}
By analogy with a nonrelativistic case we expect that in the classical limit
$\hbar\rightarrow0$ and in the unitary gauge $S=0$ the stochastically deformed system
\eqref{eq:action rKK} will describe a relativistic particle with the equations of motion
\begin{equation}\label{rel stoch eq 2}
    \dot{y}^\mu=m^{-1}g^{\mu\nu}(x,y)A_\nu(x,y)\sqrt{\dot{x}^2},\qquad
m\frac{\dot{x}^\mu}{\sqrt{\dot{x}^2}}=y^\mu.
\end{equation}
Hence in the unitary gauge and in the classical limit $m^{-1}g^{\mu\nu}A_\nu$ is expected
to be a systematic force acting on the particle. As in the previously considered model a naive introduction of the noise to the RHS of the first equation in \eqref{rel stoch eq 2} spoils the reparameterization invariance. To preserve it we have to demand from the noise correlators
\begin{equation}\label{ward ident}
    y^\mu(\tau)\lan\nu_\mu(\tau)\nu_{\mu_1}(\tau_1)\ldots\nu_{\mu_n}(\tau_n)\ran=0,
\end{equation}
where $\nu_\mu(\tau)$ is the noise. The identities \eqref{ward ident} can be viewed as the Ward-Takahashi identities for this model (see a detailed discussion in \cite{JoHu,JoHuI}). By the use of the stochastic deformation procedure we shall find below the general form of the Fokker-Planck equation associated with \eqref{rel stoch eq 2} respecting the reparameterization invariance in the case of the Gaussian white noise. A generalization of the obtained equation to the case of an arbitrary noise distribution law will be obvious.

To deform the model \eqref{eq:action rKK} we naturally realize the Heisenberg-Weyl
algebra in the linear space $V$ of smooth functions $\psi(x,y)$:
\begin{equation}
    \hat{x}^\mu=x^\mu,\qquad\hat{p}_\mu=-\hbar\partial^x_\mu,\qquad\hat{y}^\mu=y^\mu,\qquad\hat{\pi}_\mu=-\hbar\partial^y_\mu.
\end{equation}
Then the constraints \eqref{eq:action rKK} are the Weyl-ordered operators
\begin{equation}
    \hat{T}_1=\frac18(\hat{\Pi}_\mu\hat{\Pi}_\nu g^{\mu\nu}+2\hat{\Pi}_\mu
g^{\mu\nu}\hat{\Pi}_\nu+g^{\mu\nu}\hat{\Pi}_\mu\hat{\Pi}_\nu)+y^\mu\hat{p}_\mu+\phi,\qquad\hat{T}_2=y^2-m^2,
\end{equation}
where $\hat{\Pi}_\mu=\hat{\pi}_\mu+A_\mu$. These constraints are in the Abelian
involution as well\footnote{There are simpler forms of the first constraint which
preserve the classical algebra of constraints. For example, one can choose
\[
    \hat{T}_1=\frac12\hat{\Pi}_\mu g^{\mu\nu}\hat{\Pi}_\nu+y^\mu\hat{p}_\mu+\phi.
\]
Nevertheless this arbitrariness is unimportant and affects the definition of $\phi$ only
\eqref{eq:rKK}.}. The physical pure states of the system are singled out by the
conditions
\begin{equation}\label{eq:rKK phys st}
    y_0^{-1}\hat{T}_1|\psi\ran=\lan\vf|y_0^{-1}\hat{T}_1=0,\qquad\hat{T}_2|\psi\ran=\lan\vf|\hat{T}_2=0,
\end{equation}
where the inner product is defined in a standard way and we resolve the first constraint
with respect to $\hat{p}_0$. Equations \eqref{eq:rKK phys st} are invariant under the
gauge transformations
\begin{equation}
    \vf\rightarrow e^{\frac\xi\hbar}\vf,\qquad\psi\rightarrow
e^{-\frac\xi\hbar}\psi,\qquad
    A_\mu\rightarrow
    A_\mu-\partial_\mu^y\xi,\qquad
    \phi\rightarrow\phi-y^\mu\partial_\mu^x\xi.
\end{equation}
Besides the transformation of the gauge fields $A_\mu$ of the form
\begin{equation}
    A_\mu(x,y)\rightarrow A_\mu(x,y)+y_\mu\zeta(x,y),
\end{equation}
leaves Eqs. \eqref{eq:rKK phys st} unchanged. Therefore the fields $A_\mu$ have $d-1$
independent components and in the massive case one can make them orthogonal to $y^\mu$.
Introducing the phase
\begin{equation}\label{eq:phase rel}
    \vf(x,y)=e^{\frac1\hbar S(x,y)},\qquad\psi(x,y)=y^0\tilde{\rho}(x,y)e^{-\frac1\hbar
S(x,y)},
\end{equation}
we reduce the first pair of equations (the Schr\"{o}dinger equations) in formula
\eqref{eq:rKK phys st} to the system
\begin{equation}\label{eq:rKK}
\begin{split}
    &\partial_\mu^y\left[\frac\hbar2g^{\mu\nu}\partial^y_\nu\tilde\rho-g^{\mu\nu}(\partial^y_\nu
S+A_\nu)\tilde\rho\right]-\partial_\mu^x(y^\mu\tilde\rho)=0,\\
    &\frac12g^{\mu\nu}(\partial_\mu^yS+A_\mu)(\partial_\nu^yS+A_\nu)+y^\mu\partial_\mu^xS+\phi=-\frac\hbar2\partial_\mu^y[g^{\mu\nu}(\partial_\nu^yS+A_\nu)]-\frac{\hbar^2}8\partial_{\mu\nu}^yg^{\mu\nu}.
\end{split}
\end{equation}
The first equation is the conservation law of the probability density current whose time
component is the probability density function. The second equation is the definition of
$\phi$.

The second pair of equations in \eqref{eq:rKK phys st} is taken into account by the
equalities
\begin{equation}\label{eq:equali}
    \tilde\rho=\de(y^2-m^2)\bar\rho(x,y^i),\qquad S=\ln\de(y^2-m^2)+\bar{S}(x,y^i).
\end{equation}
On substituting these equalities to the first equation in \eqref{eq:rKK} we obtain a
relativistic generalization of the Klein-Kramers equation
\cite{LandLif,DebMalRiv,DunHan,Fa,AngFra,ChAcKr,CCPDTH}\footnote{The position of the tensor $g^{ij}$ in Eq. \eqref{eq:rKK lab} between two derivatives $\partial_i^y$ is uniquely determined by the requirement that the evolution keeps the mass-shell condition. It is that order which results in the J\"{u}ttner stationary distribution derived from the relativistic Klein-Kramers equation in the papers \cite{DunHan}.}
\begin{equation}\label{eq:rKK lab}
    \partial_i^y\left[\frac{g^{ij}}{y^0}\left(\frac\hbar2\partial_j^y\bar\rho-(\partial_j^y\bar
S+A_j)\bar\rho\right)\right]-\partial_\mu^x\left(\bar\rho\frac{y^\mu}{y^0}\right)=0,
\end{equation}
where the positive root of the mass-shell condition is chosen: $y^0=\sqrt{m^2+(y_i)^2}$. The logarithm of the $\de$-function in \eqref{eq:equali} does not actually contribute to Eqs. \eqref{eq:rKK lab} because of the orthogonality of the tensor $g^{\mu\nu}$ to $y_\mu$. Now it is not difficult to construct a path-integral representation of the transition
probability and the corresponding Langevin equations \eqref{eq:langevin}. For brevity we
give the Langevin equations only
\begin{equation}\label{eq:langevin KK rel}
\begin{gathered}
    \dot{x}^\mu=\frac{y^\mu}{y^0},\qquad\dot{y}^i=\frac{g^{ij}}{y^0}(\partial_j^y\bar
S+A_j)+\frac\hbar2\partial_j^y\left(\frac{g^{ij}}{y^0}\right)+\nu^i,\\
    \lan\nu^i(\tau)\ran=0,\qquad\lan\nu^i(\tau)\nu^j(\tau')\ran=\hbar\frac{g^{ij}}{y^0}\de(\tau-\tau'),
\end{gathered}
\end{equation}
where $\nu^i$ is a Gaussian white noise and we assume that $g^{ij}(x,y)$ is positively
definite. Recall that the stochastic equations \eqref{eq:langevin KK rel} are understood
in the Ito sense. Thus we confirm our expectations that equation \eqref{eq:rKK lab} or
\eqref{eq:langevin KK rel} describes a relativistic particle in the gauge of a laboratory
time \eqref{eq:temp gauge} under the influence of noise.

As far as the tensor $g^{\mu\nu}$ is concerned its physical meaning is comprehended from
the Langevin equations \eqref{eq:langevin KK rel}. We remark only two possible choices
\begin{equation}
    g_1^{\mu\nu}=\frac{y^\mu y^\nu}{y^2}-\eta^{\mu\nu},\qquad
g_2^{\mu\nu}=-(\eta^\mu_\rho-\frac{n^\mu y_\rho}{n_\s
y^\s})(\eta^{\rho\nu}-\frac{y^\rho n^\nu}{n_\la y^\la}),\quad n^2=1.
\end{equation}
The first tensor implies that the effect of the stochastic force on a relativistic
particle is isotropic in a momentary comoving frame. The second tensor corresponds to the
isotropic influence of the stochastic force in the frame distinguished by the vector
$n^\mu$. It is interesting to note that in the first case, which is minimal from the
mathematical point of view since new objects are not introduced, the tensor $g^{ij}_1$
can be interpreted as the inverse metric in the momentum space with the coordinates
$y^i$. Keeping in mind that
\begin{equation}
    (\det{g_1^{ij}})^{1/2}=m^{-1}y^0,
\end{equation}
we can rewrite Eq. \eqref{eq:rKK lab} in a covariant way. The metric $g^1_{ij}$ is a
metric of a constant positive curvature and, consequently, it is conformally flat. The
second tensor $g^{ij}_2$ becomes Euclidean provided that $n^\mu=\de^\mu_0$.

To conclude this example we briefly discuss a representation of the transition
probability in terms of the proper time. The main observation allowing to formulate Eqs.
\eqref{eq:langevin KK rel} in the proper time gauge is the following. The retarded Green
function of the twisted forward Schr\"{o}dinger equation \eqref{eq:rKK phys st}, that is
the transition probability, is the kernel of
\begin{equation}\label{eq:trans prob rKK}
    -\frac1{\vf
y_0^{-1}\hat{T}_1\vf^{-1}}=-y^0\frac{1}{\tilde{\vf}\hat{T}_1\tilde{\vf}^{-1}}=\int\limits_0^\infty
d\tau \frac{y^0}{\hbar m} e^{\frac{\tau}{\hbar
m}\tilde{\vf}\hat{T}_1\tilde{\vf}^{-1}},
\end{equation}
where $\tilde\vf:=y_0^{-1}\vf$ and the last equality is proved by the descent method as
it was done in considering a relativistic generalization of the Fokker-Planck equation.
The operator in the exponent is proportional to the operator acting on $\tilde\rho$ in
the first equation in formula \eqref{eq:rKK}.

The kernel
\begin{equation}\label{eq:trans prob rKK prop}
    \lan x,y| \exp{\left[\frac{\tau}{\hbar
m}\tilde{\vf}\hat{T}_1\tilde{\vf}^{-1}\right]}|x',y'\ran
\end{equation}
is the probability density of particle's arrival to the point $(x,y)$ from the point
$(x',y')$ with the proper time $\tau$ measured in the momentary comoving frame. The
factor $y^0m^{-1}$ in \eqref{eq:trans prob rKK} is caused by the passage from the
momentary comoving frame to the lab frame. The transition probability possesses the
Markov property with respect to the proper time $\tau$. It conserves the normalization of
the probability density function, i.e., the probability to find a particle with the
proper time $\tau$ in some point of the space-time with a certain momentum is equal to
unity. Since the generator of evolution commutes with the operator $\hat{T}_2$ the
transition probability \eqref{eq:trans prob rKK prop} respects the mass-shell condition.

The Langevin equations associated with \eqref{eq:trans prob rKK prop} read as
\begin{equation}
\begin{gathered}
    m\dot{x}^\mu=y^\mu,\qquad \dot{y}^\mu=m^{-1}g^{\mu\nu}(\partial_\nu^y S+A_\nu)+\hbar
m^{-1}\partial_\nu^yg^{\mu\nu}/2+\nu^\mu,\\
    \lan\nu^\mu(\tau)\ran=0,\qquad\lan\nu^\mu(\tau)\nu^\nu(\tau')\ran=\hbar m^{-1}
g^{\mu\nu}\de(\tau-\tau'),
\end{gathered}
\end{equation}
whence we infer that $\tau$ is indeed the proper time. As before the noise $\nu^\mu$ is a
Gaussian white noise. The representation \eqref{eq:trans prob rKK} of the transition
probability is especially useful when a relativistic particle has a finite lifetime. In
that case the integral \eqref{eq:trans prob rKK} over $\tau$ is cut on the upper limit by
the lifetime of the particle provided, of course, that the particle was created at the
initial moment.

\subsubsection{Stochastic Lorentz-Dirac equation}

A relativistic generalization of the Klein-Kramers equation regarded in the preceding
example does not apply to a charged particle with a radiation reaction taken into
account. If we apply the above scheme to the so-called Landau-Lifshitz equation
\cite{LandLifsh}, which is obtained from the Lorentz-Dirac equation by the reduction of
order procedure, then the vanishing external systematic force entails a zero dissipation
force in the resulting stochastic equation. That is why we should stochastically deform
the Lorentz-Dirac equation itself. The procedure of stochastic deformation is very similar
to what we have considered in the previous example. Therefore we mention the crucial
points only.

We start from the Hamiltonian action of the form
\begin{multline}\label{eq:action sLD}
    S_H[x,\bar{x},y,\bar{y},w,\bar{w},\la,\mu,\ka]=\int
d\tau(\bar{x}_\mu\dot{x}^\mu+\bar{y}_\mu\dot{y}^\mu+\bar{w}_\mu\dot{w}^\mu
    -\la T_1-\mu T_2-\ka T_3),\\
    T_1:=\frac12g^{\mu\nu}\bar{W}_\mu\bar{W}_\nu-w^2y^\mu\bar{W}_\mu+
    w^\mu\bar{y}_\mu+y^\mu\bar{x}_\mu+\phi,\qquad
    T_2:=y^2-1,\qquad T_3:=yw,
\end{multline}
where $\bar{W}_\mu:=\bar{w}_\mu+A_\mu$ is a covariant momentum, $A_\mu(x,y,w)$ and
$\phi(x,y,w)$ are gauge fields and $g^{\mu\nu}(x,y,w)$ is a symmetric Lorentz tensor
orthogonal to $y_\mu$. For simplicity we assume that the particle has a unit mass. The
constraints $T_1$, $T_2$ and $T_3$ are of the first class with the
algebra
\begin{equation}\label{eq:constr alg class sLD}
    \{T_1,T_2\}=-2T_3,\qquad\{T_1,T_3\}=w^2T_2.
\end{equation}

A relevant part of the equations of motion resulting from the action \eqref{eq:action
sLD} reads as
\begin{equation}
    \dot{x}^\mu=\la y^\mu,\qquad\dot{y}^\mu=\la
w^\mu,\qquad\dot{w}^\mu=\la(g^{\mu\nu}\bar{W}_\nu-w^2y^\mu),\qquad y^2=1,\qquad
yw=0.
\end{equation}
These equations suggest that in the classical limit and in the unitary gauge the
stochastically deformed model \eqref{eq:action sLD} describes a relativistic particle
obeying the equations of motion
\begin{equation}\label{eq:class sLD}
    \dddot{x}^\mu+\ddot{x}^2\dot{x}^\mu=g^{\mu\nu}A_\nu,\qquad \dot{x}^\mu=y^\mu,\qquad
\dot{y}^\mu=w^\mu,
\end{equation}
where the derivatives are taken with respect to the proper time. The LHS of the first
equation is proportional to the Lorentz-Dirac force and we refer to this equation as the
Lorentz-Dirac equation. The mass term is contained in the RHS of this equation.

The Heisenberg-Weyl algebra is naturally realized in the linear space $V$ of smooth
functions $\psi(x,y,w)$:
\begin{equation}
    \hat{x}^\mu=x^\mu,\qquad\hat{\bar{x}}_\mu=-\hbar\partial^x_\mu,\qquad\hat{y}^\mu=y^\mu,\qquad\hat{\bar{y}}_\mu=-\hbar\partial^y_\mu,\qquad\hat{w}^\mu=w^\mu,\qquad\hat{\bar{w}}_\mu=-\hbar\partial^w_\mu.
\end{equation}
The constraints turn into the appropriate Weyl-ordered operators. Their algebra coincides
with the classical algebra of constraints \eqref{eq:constr alg class sLD}. The physical
states of the stochastic system are specified by
\begin{equation}\label{eq:sLD phys st}
    y_0^{-1}\hat{T}_1|\psi\ran=\lan\vf|y_0^{-1}\hat{T}_1=0,\qquad\hat{T}_2|\psi\ran=\lan\vf|\hat{T}_2=0,\qquad\hat{T}_3|\psi\ran=\lan\vf|\hat{T}_3=0,
\end{equation}
where the standard inner product is understood. The first pair of equations in formula
\eqref{eq:sLD phys st} is just the forward and backward Schr\"{o}dinger equations.

Making a substitution of the form \eqref{eq:phase rel} into the first equation in
\eqref{eq:sLD phys st} we arrive at\footnote{The proper time Liouville equation for the Lorentz-Dirac equation can be found, for example, in \cite{Hak}.}
\begin{multline}\label{eq:sLD}
    \partial_\mu^w\left[\frac\hbar2g^{\mu\nu}\partial^w_\nu\tilde\rho-\left(g^{\mu\nu}(\partial^w_\nu
S+A_\nu)-w^2y^\mu\right)\tilde\rho\right]-\partial_\mu^y(w^\mu\tilde\rho)-\partial_\mu^x(y^\mu\tilde\rho)=0,\\
    \shoveleft{\frac12g^{\mu\nu}(\partial_\mu^wS+A_\mu)(\partial_\nu^wS+A_\nu)-w^2y^\mu\partial_\mu^wS+w^\mu\partial_\mu^yS+y^\mu\partial_\mu^xS-yw+\phi=}\\
    -\frac\hbar2\partial_\mu^w[g^{\mu\nu}(\partial_\nu^wS+A_\nu)]-\frac{\hbar^2}8\partial_{\mu\nu}^wg^{\mu\nu}.
\end{multline}
As usual the first equation is the conservation law of the probability density current.
The probability density function is the time component of this current.
The rest of equations defining the physical state results in
\begin{equation}
    \tilde\rho=\de(y^2-1)\de(yw)\bar\rho(x,y^i,w^i),\qquad
S=\ln\left[\de(y^2-1)\de(yw)\right]+\bar{S}(x,y^i,w^i).
\end{equation}
Under the above conditions the first equation in \eqref{eq:sLD} can be reduced after a
little algebra to
\begin{equation}\label{eq:sLD lab}
    \partial_i^w\left[\left(\frac\hbar2g^{ij}\partial^w_j-\left(g^{ij}(\partial_j^w\bar
S+A_j)-w^2y^i\right)\right)\frac{\bar\rho}{y_0^2}\right]-\partial_i^y\left(w^i\frac{\bar\rho}{y_0^2}\right)-\partial_\mu^x\left(y^\mu\frac{\bar\rho}{y_0^2}\right)=0,
\end{equation}
where we have assigned $y^0=\sqrt{1+(y_i)^2}$ and $w^0=y_0^{-1}y_iw_i$. Recall that the
probability density function is $y_0^{-1}\bar\rho$. Now a path-integral representation of
the transition probability associated with Eq. \eqref{eq:sLD lab} is straightforward. The
Langevin equations \eqref{eq:langevin} with the Ito prescription become
\begin{equation}\label{eq:langevin sLD}
\begin{gathered}
    \dot{x}^\mu=\frac{y^\mu}{y^0},\qquad\dot{y}^i=\frac{w^i}{y^0},\qquad\dot{w}^i=\frac{g^{ij}}{y^0}(\partial_j^w\bar
S+A_j)-w^2\frac{y^i}{y^0}+\frac\hbar2\partial_j^w\left(\frac{g^{ij}}{y^0}\right)+\nu^i,\\
    \lan\nu^i(\tau)\ran=0,\qquad\lan\nu^i(\tau)\nu^j(\tau')\ran=\hbar\frac{g^{ij}}{y^0}\de(\tau-\tau'),
\end{gathered}
\end{equation}
where $\nu^i$ is a Gaussian white noise and $g^{ij}(x,y,w)$ is positively definite. As
regards explicit forms of the tensor $g^{ij}$ see the discussion in the previous
example\footnote{In the series of papers \cite{JoHu,JoHuI} the stochastic Lorentz-Dirac
equation was derived from quantum electrodynamics in the paradigm of the Feynman-Vernon
influence functional. There was obtained an explicit expression for the tensor
$g^{\mu\nu}$ that is a nonlocal in $x^\mu$ operator orthogonal to $y_\mu$. A
generalization of Eqs. \eqref{eq:sLD} and \eqref{eq:sLD lab} to this case is
straightforward.}. The obtained stochastic equations are equivalent to the Lorentz-Dirac
equation \eqref{eq:class sLD} with a random force which is rewritten in the gauge of a
laboratory time, that is equation \eqref{eq:langevin sLD} is the stochastic Lorentz-Dirac
equation.

The proper time representation of the transition probability from the point $(x',y',w')$
to the point $(x,y,w)$ takes the form
\begin{equation}
    \lan x,y,w|\int\limits_0^\infty d\tau \hbar^{-1}y^0 \exp{\left[\frac{\tau}{\hbar
}\tilde{\vf}\hat{T}_1\tilde{\vf}^{-1}\right]}|x',y',w'\ran,
\end{equation}
where $\tilde\vf:=y_0^{-1}\vf$ and $\tilde{\vf}\hat{T}_1\tilde{\vf}^{-1}$ is the operator
acting on $\tilde\rho$ in the first equation of \eqref{eq:sLD}. Besides, the initial
state should satisfy the mass-shell condition and its differential consequence.
Therefore the proper time stochastic Lorentz-Dirac equation is the following
\begin{equation}
\begin{gathered}
    \dot{x}^\mu=y^\mu,\qquad\dot{y}^\mu=w^\mu,\qquad
\dot{w}^\mu=g^{\mu\nu}(\partial_\nu^w
S+A_\nu)-w^2y^\mu+\frac\hbar{2}\partial_\nu^wg^{\mu\nu}+\nu^\mu,\\
    \lan\nu^\mu(\tau)\ran=0,\qquad\lan\nu^\mu(\tau)\nu^\nu(\tau')\ran=\hbar
g^{\mu\nu}\de(\tau-\tau'),
\end{gathered}
\end{equation}
where $\nu^\mu$ is a Gaussian white noise.

\subsection{Free relativistic field models}\label{rel fields}

In this subsection we consider the stochastic deformation of two free models: a
relativistic real scalar field and an electromagnetic field. We regard the scalar field
model as a prototype of any field model. Besides, it is the simplest model of a
one-dimensional crystal in the continuum limit, where the scalar field describes the
displacements of points of a medium from their equilibrium positions. The stochastically
deformed model of Maxwell's fields allows us to touch the problem of electromagnetic
fluctuations \cite{LandLifstat} bringing it in accordance with the general scheme
advocated in this paper.

\subsubsection{Scalar field}

Consider a real scalar field $\phi(x)$ on the Minkowski space $\mathbb{R}^{1,d-1}$
with the dynamics governed by the action
\begin{equation}\label{eq:action sc f}
    S[\phi]=\frac12\int{d^dx(\partial_\mu\phi\partial^\mu\phi-m^2\phi^2)},
\end{equation}
where $m$ is a mass. The corresponding Hamiltonian action has the form
\begin{equation}
    S_H[\phi,\pi]=\int{d^dx\left[\pi\dot{\phi}-\frac12(\pi^2+\partial_i\phi\partial_i\phi+m^2\phi^2)\right]},
\end{equation}
where the dot denotes the derivative with respect to time. The Hamilton equations are
\begin{equation}\label{eq:ham eq sc}
    \dot{\phi}=\pi,\qquad\dot{\pi}=\Delta\phi-m^2\phi.
\end{equation}
The Noether theorem gives the conserved $4$-momentum
\begin{equation}\label{eq:transl gen}
    \mathcal{P}_0\equiv
H=\frac12\int{d\mathbf{x}(\pi^2+\partial_i\phi\partial_i\phi+m^2\phi^2)
},\qquad\mathcal{P}_i=\int d\mathbf{x}\pi\partial_i\phi,
\end{equation}
which is the generator of translations.

Now we are going to define the stochastic deformation. The canonical coordinates $\phi(\mathbf{x})$
and canonical momenta $\pi(\mathbf{x})$ turn into the generators of the Heisenberg-Weyl
algebra
\begin{equation}\label{eq:comm rel field}
    [\hat{\phi}(\mathbf{x}),\hat{\pi}(\mathbf{y})]=\hbar\de(\mathbf{x}-\mathbf{y}).
\end{equation}
Since we deform the linear model its Heisenberg equations coincide with the Hamilton
equations. The Heisenberg equations can be formally solved by means of the expansion in
terms of the complete set of solutions of the classical equations of motion in the form
of plane waves. The field operator and its canonical conjugate read as
\begin{equation}\label{eq:heis sol sc}
\begin{split}
    \hat{\phi}(x)&=\int\frac{d\spp}{(2\pi)^{d-1}}\frac1{\sqrt{p_0}}\left[\hat{a}(\spp)\cos(p_\mu
x^\mu)+\hat{b}(\spp)\sin(p_\mu x^\mu)\right],\\
    \hat{\pi}(x)&=\int\frac{d\spp}{(2\pi)^{d-1}}\sqrt{p_0}\left[\hat{b}(\spp)\cos(p_\mu
x^\mu)-\hat{a}(\spp)\sin(p_\mu x^\mu)\right],
\end{split}
\end{equation}
where $p_0:=\sqrt{m^2+\spp^2}$. As it follows from \eqref{eq:comm rel field} the
operators $\hat{a}(\spp)$ and $\hat{b}(\spp)$ obey the commutation relations
\begin{equation}
    [\hat{a}(\spp),\hat{b}(\spk)]=\hbar(2\pi)^{d-1}\de(\spp-\spk).
\end{equation}
The Weyl-ordered operators of the generators of translations \eqref{eq:transl gen} look
like
\begin{equation}\label{eq:transl gen q}
    \hat{\mathcal{P}}_\mu=\int\frac{d\spp
}{(2\pi)^{d-1}}\frac{p_\mu}2[\hat{a}^2(\spp)+\hat{b}^2(\spp)].
\end{equation}
The solution \eqref{eq:heis sol sc} also implies that
\begin{equation}\label{eq:comm rel ff field}
    [\hat{\phi}(x),\hat{\phi}(y)]=\hbar\int\frac{d\spp \sin[p_\mu(y-x)^\mu]}{(2\pi)^{d-1}
p_0}=-2\hbar\sign(x^0-y^0)\bar{G}(x-y),
\end{equation}
where $\bar{G}(x-y)$ is the symmetric Green function (see, e.g., \cite{BjoDreII})
\begin{equation}
    (\Box+m^2)\bar{G}(x)=\de^d(x),\qquad\bar{G}(x-y)=\bar{G}(y-x).
\end{equation}
As in the case of quantum field theory two field operators separated by a space-like
interval commute.

The key ingredient of stochastic field theory is the propagator, which we define in a
standard way
\begin{equation}\label{eq:propag}
    G(x,y):=\lan T\{\hat{\phi}(x)\hat{\phi}(y)\}\ran,
\end{equation}
where $T$ is a chronological ordering. Any other correlator can be expressed in terms of
the propagator by means of the Wick theorem and the commutation relation \eqref{eq:comm
rel ff field}.

Certainly, an explicit form of the propagator depends on the state entering in its
definition. Nevertheless the propagator possesses the properties following from the rule
of differentiation of the chronological ordering and the Heisenberg equations
\begin{equation}
    (\Box_x+m^2)G(x,y)=-\hbar\de^d(x-y),\qquad G(x,y)=G(y,x),
\end{equation}
regardless of the state which does not depend on time explicitly. Therefore the
propagator can be represented as
\begin{equation}\label{eq:propag 1}
    G(x,y)=-\hbar\bar{G}(x-y)+\ldots,
\end{equation}
where dots denote some solution of the Klein-Gordon equation.

Moreover if the state $\hat\rho$ is translation invariant
\begin{equation}
    [\hat{\rho},\hat{\mathcal{P}}_\mu]=0,
\end{equation}
then the obvious identity
\begin{equation}
    \Sp\left([\hat{\mathcal{P}}_\mu,\hat\rho T\{\hat{\phi}(x)\hat{\phi}(y)\}]\right)=0
\end{equation}
entails a translational invariance of the propagator, viz.
\begin{equation}
    G(x,y)=G(x-y).
\end{equation}
Analogously it can be proved that an invariance of the state with respect to the Lorentz
transformations results in a dependence of the propagator on the space-time interval
only.

Unlike quantum field theory there is no distinguished ground state in our case since the
Hamiltonian \eqref{eq:transl gen q} has an unbounded spectrum. That is why we define the
state $\hat\rho$ in such a way that the propagator \eqref{eq:propag} becomes minimal in
view of the above considerations, that is the dotted terms in formula
\eqref{eq:propag 1} vanish.

Let us achieve this goal gradually. Firstly, suppose that $\hat\rho$ commutes with the
Hamiltonian, i.e., the state is invariant with respect to translations in time. Then from
the equality
\begin{equation}
    \lan[\hat{a}(\spp)\hat{b}(\spk),\hat{H}]\ran=p_0\lan\hat{b}(\spp)\hat{b}(\spk)\ran-k_0\lan\hat{a}(\spp)\hat{a}(\spk)\ran=0,
\end{equation}
and the same equality with $\spp$ and $\spk$ interchanged we find
\begin{equation}
    \lan\hat{a}(\spp)\hat{a}(\spk)\ran=\lan\hat{b}(\spp)\hat{b}(\spk)\ran=f(\spp,\spk)\de(p_0-k_0),\qquad
f(\spp,\spk)=f(\spk,\spp).
\end{equation}
Furthermore, from the equalities
\begin{equation}
\begin{split}
    \lan[\hat{a}(\spp)\hat{a}(\spk),\hat{H}]\ran&=\frac12\lan k_0
[\hat{a}(\spp)\hat{b}(\spk)+\hat{b}(\spk)\hat{a}(\spp)]+p_0[\hat{a}(\spk)\hat{b}(\spp)
+\hat{b}(\spp)\hat{a}(\spk)]\ran=0,\\
    \lan[\hat{b}(\spp)\hat{b}(\spk),\hat{H}]\ran&=-\frac12\lan k_0
[\hat{b}(\spp)\hat{a}(\spk)+\hat{a}(\spk)\hat{b}(\spp)]+p_0[\hat{b}(\spk)\hat{a}(\spp)
+\hat{a}(\spp)\hat{b}(\spk)]\ran=0,
\end{split}
\end{equation}
we infer that
\begin{equation}
    \frac12\lan\hat{a}(\spp)\hat{b}(\spk)+\hat{b}(\spk)\hat{a}(\spp)\ran=g(\spp,\spk)\de(p_0-k_0),\qquad
g(\spp,\spk)=-g(\spk,\spp).
\end{equation}

By the same way if we assume that the state $\hat\rho$ is also invariant with respect to the
spatial translations then
\begin{equation}
    \lan\hat{a}(\spp)\hat{a}(\spk)\ran=\lan\hat{b}(\spp)\hat{b}(\spk)\ran=f(p)\de(\spp-\spk),\qquad\frac12\lan\hat{a}(\spp)\hat{b}(\spk)+\hat{b}(\spk)\hat{a}(\spp)\ran=0.
\end{equation}
Thus we have for the dotted terms in formula \eqref{eq:propag 1}
\begin{equation}\label{eq:dotted int}
    \int\frac{d\spp}{(2\pi)^{2d-2}}\frac{f(p)}{p_0}\cos[p_\mu(x-y)^\mu],
\end{equation}
where one should bear in mind that the propagator expressed in terms of the operators
$\hat{a}$ and $\hat{b}$ looks like
\begin{multline}
    G(x,y)=\\
    \int\frac{d\spp d\spk}{(2\pi)^{2d-2}\sqrt{p_0k_0}} \lan\left[\hat{a}(\spp)\cos(p_\mu
x^\mu)+\hat{b}(\spp)\sin(p_\mu x^\mu)\right]\left[\hat{a}(\spk)\cos(k_\mu
y^\mu)+\hat{b}(\spk)\sin(k_\mu y^\mu)\right]\ran,
\end{multline}
provided $x^0>y^0$. The additional requirement of an invariance of the state with respect
to the Lorentz group yields that the function $f(p)$ contributes to the integral
\eqref{eq:dotted int} as a constant. Consequently, this integral is proportional to the
so-called Hadamard function \cite{BjoDreII}
\begin{equation}
    G^{(1)}(x-y):=\int\frac{d\spp\cos[p_\mu(x-y)^\mu]}{(2\pi)^{d-1}p_0}.
\end{equation}
The Hadamard function does not vanish for the points $x$ and $y$ separated by a spatial
interval. Hence if we want to obtain a causal theory we have to set the proportionality
constant to zero that leaves us with the propagator proportional to the symmetric Green
function.

The above considerations prove that the propagator proportional to the symmetric Green
function is unique for the stochastically deformed model \eqref{eq:action sc f} under the
causality condition and the requirement of the Poincar\'{e} invariance of the state.
Explicitly such a pure state $\hat\rho=|\Psi\ran\lan\Phi|$ can be specified as
\begin{equation}\label{eq:ground st}
    |\Psi\ran\lan\Phi|=\prod\limits_{\spp}|\Psi_\spp\ran\lan\Phi_\spp|,
\end{equation}
where
\begin{equation}
    [\hat{a}^2(\spp)+\hat{b}^2(\spp)]|\Psi_\spp\ran=\lan\Phi_\spp|[\hat{a}^2(\spp)+\hat{b}^2(\spp)]=0.
\end{equation}
In particular this implies
\begin{equation}
    \hat{\mathcal{P}}_\mu|\Psi\ran=\lan\Phi|\hat{\mathcal{P}}_\mu=0.
\end{equation}

The propagator \eqref{eq:propag} naturally appears when one computes, for example, the
transition probability from the state $|\Psi_1\ran\lan\Phi_1|$ to the state
$|\Psi_2\ran\lan\Phi_2|$
\begin{equation}
    \lan\Phi_2|\hat{U}|\Psi_1\ran\lan\Phi_1|\hat{U}^{-1}|\Psi_2\ran,
\end{equation}
where $\hat{U}$ is the evolution operator over an infinite time and the states
$|\Psi_1\ran\lan\Phi_1|$ and $|\Psi_2\ran\lan\Phi_2|$ are obtained from the ground state
\eqref{eq:ground st} by the action of the operators $\hat{a}$ and $\hat{b}$. The
perturbation techniques is in a perfect analogy with such a procedure in quantum field
theory. In our case the so-called closed-time-path formalism (see, e.g., \cite{JoHu,LandWee} and
references therein) seems to be especially useful for calculations. If the classical
current is introduced into the model \eqref{eq:action sc f} then the solution
\eqref{eq:heis sol sc} of the Heisenberg equations is the general solution of the
homogeneous equation only. A particular solution of the inhomogeneous equation with
current is given by a convolution of the retarded Green function with this classical
current. In that case it is the convolution of the retarded Green function with
current that is the average field over the ground state \eqref{eq:ground st}.

\subsubsection{Electromagnetic field}

As usual we start from the classical Maxwell action\footnote{We follow \cite{HeTe} in the
deformation procedure for the model of free electromagnetic field.}
\begin{equation}\label{eq:action Max}
    S[A]=-\frac14\int d^4xF_{\mu\nu}F^{\mu\nu}=\frac12\int
d^4x\left[\dot{A}_i^2-2\partial_iA_0\dot{A}_i+(\partial_iA_0)^2-(\partial_iA_j)^2+\partial_iA_j\partial_jA_i\right],
\end{equation}
where $F=dA$ is the strength tensor of the electromagnetic field. Making the Legendre
transformations on $\dot{A}_i$ only and introducing the canonical momenta
\begin{equation}
    \pi_i=\frac{\partial L}{\partial\dot{A}^i}=\partial_iA_0-\dot{A}_i,
\end{equation}
we arrive at the Hamiltonian action
\begin{equation}
    S_H=\int
d^4x\left\{\pi_i\dot{A}^i-\frac12\left[\pi_i^2+(\partial_iA_j)^2-\partial_iA_j\partial_jA_i\right]+A_0\partial_i\pi^i\right\}.
\end{equation}
As we see the time component of the gauge fields $A_\mu$ is merely the Lagrange
multiplier to the Gauss law constraint. The expression in square brackets with $1/2$ is
the density of the physical Hamilton function $H_0$. Thus we have a model with
first class constraints.

According to the standard BFV-quantization scheme for the first class constrained models we introduce the canonically conjugate ghost pairs $(c,P)$ and $(\bar c,\bar P)$
\begin{equation}
    [\hat{c}(\spx),\hat{P}(\spy)]=[\hat{\bar{c}}(\spx),\hat{\bar{P}}(\spy)]=\hbar\de(\spx-\spy),\qquad\gh
c=\gh\bar{c}=-\gh P=-\gh\bar{P}=1,
\end{equation}
where square brackets denote graded commutators. Besides we introduce the canonical
momentum $\pi_0$, $\gh\pi_0=0$, to the Lagrange multiplier $A_0$. Then we construct  the
BRST-charge
\begin{equation}
    \hat\Omega=\int
d\spx(\hat{c}\partial_i\hat\pi_i+\hat{\bar{c}}\hat\pi_0),\qquad[\hat\Omega,\hat\Omega]=2\hat\Omega^2=0,\quad\gh\hat\Omega=1.
\end{equation}
Physical observables and the Hamiltonian should commute with the BRST-charge. In our case
the Hamiltonian $\hat{H}_0$ obviously satisfies this requirement.
The gauge fixing fermion is taken in the form
\begin{equation}
    \hat\psi=\int
d\spx\left[\hat{P}\hat{A}_0+\hat{\bar{P}}\left(\partial_i\hat{A}_i-\frac\al2\hat{\pi}_0\right)\right],\qquad\gh\hat\psi=-1,
\end{equation}
where $\al$ is an arbitrary constant.

Therefore the gauge fixed Hamiltonian looks like
\begin{multline}
    \hat{H}:=\hat{H}_0+[\hat\Omega,\hat\psi]=\\
    \int
d\spx\left\{\frac12\left[\hat{\pi}_i^2+(\partial_i\hat{A}_j)^2-\partial_i\hat{A}_j\partial_j\hat{A}_i\right]-\hat{A}_0\partial_i\hat{\pi}^i-\hat{\pi}_0\left(\partial_i
\hat{A}^i+\frac\al2\hat{\pi}_0\right)-\hat{\bar{c}}\hat{P}+\partial_i\hat{c}\partial_i\hat{\bar{P}}\right\},
\end{multline}
and commutes with the BRST-charge as well.
The corresponding Heisenberg equations read as
\begin{equation}
\begin{aligned}
    \dot{\hat{A}}_i&=\partial_i\hat{A}_0-\hat\pi_i,&\qquad\dot{\hat{A}}_0&=\partial_i
\hat{A}_i-\al\hat\pi_0,&\qquad\dot{\hat{c}}&=\hat{\bar{c}},&\qquad\dot{\hat{P}}&=-\Delta\hat{\bar{P}},\\
    \dot{\hat{\pi}}_i&=\partial_i\partial_j\hat{A}_j-\Delta
\hat{A}_i-\partial_i\hat{\pi}_0,&\qquad\dot{\hat\pi}_0&=-\partial_i\hat\pi_i,&\qquad\dot{\hat{\bar{c}}}&=\Delta\hat{c},&\qquad\dot{\hat{\bar{P}}}&=-\hat{P}.
\end{aligned}
\end{equation}
As expected the ghosts dynamics are split of the fields dynamics and we do not enlarge on
them henceforth. Combining the Heisenberg equations it is not difficult to obtain
\begin{equation}
    \Box\hat{A}_0=(\al-1)\partial_i\hat\pi_i,\qquad\Box
\hat{A}_i=(1-\al)\partial_i\hat\pi_0,\qquad\Box\hat\pi_i=0.
\end{equation}
In the Feynman gauge $\al=1$ the Heisenberg equations on the fields reduce to the wave
equations and we can use all the results of the preceding example regarding the model of
a scalar field. The physical states are those containing solely transverse photons.

Hence in the Feynman gauge the Poincar\'{e}-invariant causal propagator of the fields
$A_\mu$ becomes
\begin{equation}\label{eq:propag em}
    G(x-y)=\lan
T\{\hat{A}_\mu(x)\hat{A}_\nu(y)\}\ran=\hbar\eta_{\mu\nu}\bar{G}(x-y),\qquad\Box\bar{G}(x)=\de^d(x),
\end{equation}
where $\bar{G}(x)$ is the symmetric Green function, what coincides with the well-known
result of \cite{LandLifstat} in the case of a transparent medium $\e=\mu=1$. The
correlators of the electromagnetic fields $F_{\mu\nu}$ are straightforwardly obtained
from \eqref{eq:propag em}.


\section{Concluding remarks}

There are, of course, many important open problems regarding the procedure advocated here. We mention a few of them only. Apart from the study both analytical and numerical of solutions of the derived relativistic equations we distinguish a more detailed investigation of the model with nonlinear phase space. In spite of the fact that we considered several models of this kind their a priori interpretation remains unclear. For instance, in studying the stochastically deformed model of a relativistic particle with Hamilton function quadratic in momenta and with a nonlinear phase space, we found after constructing a path-integral representation of the transition probability that the probability distribution function of the random force is non-Gaussian. It is desirable to obtain a simple method to discover the noise distribution from the initial classical model. Furthermore we only touched the problem of stochastic deformation of field theories. The next step is to consider interacting models and their renormalization properties. It is also interesting to investigate the structure of the gauge transformations and corresponding functional gauge fields associated with the changing of the phase $S$ in the stochastic field theory framework. Notice that the gauged formulation of stochastic mechanics developed in this paper allows us to represent it in terms of sections of vector bundles analogously to quantum mechanics. For the Schr\"{o}dinger equations of the form \eqref{shrodinger eqs nonrel} the pair of functions $\vf(x)$ and $\psi(x)$ describing a pure state of a stochastic system can be viewed as section (written in the light cone coordinates) of a vector bundle over the space-time with the typical fiber $\mathbb{R}^2$ and the structure group $SO(1,1)$. Then the states should be described by time-like or isotropic sections to guarantee the non-negativity of probability density functions. The gauge fields are connections on this vector bundle. A similar construction on the Whitney sum of tangent and cotangent bundles $TM\oplus T^*M$ is called an almost generalized product structure (see, for introduction, \cite{Zabz}) that is a real analog of an almost generalized complex structure \cite{Hitch}.

\begin{acknowledgments}

I am indebted to an anonymous referee for valuable comments. I am also grateful to A.A. Sharapov for reading of the draft of this paper, discussions and useful suggestions. I  appreciate Prof. V.G. Bagrov for fruitful discussions on various subjects of relativistic wave equations. This work was supported by the RFBR grant 06-02-17352 and the grant for Support of Russian Scientific Schools SS-871.2008.2.

\end{acknowledgments}


\end{document}